\documentclass[useAMS,usenatbib]{mn2e}

\usepackage{natbib}
\usepackage{bm}
\usepackage{psfrag}
\usepackage{mathrsfs}
\usepackage{amssymb,amsmath}
\usepackage{graphicx}
\usepackage{float}
\usepackage{epstopdf}

\newcommand{\be}{\begin{eqnarray} }
\newcommand{\ee}{ \end{eqnarray} }
\newcommand\captionof[1]{\def\@captype{#1}\caption}

\newcommand{\pasa}{PASA}
\newcommand{\apj}{ApJ}
\newcommand{\apjs}{ApJS}
\newcommand{\apjl}{ApJL}
\newcommand{\aap}{A{\&}A}

\newcommand{\mnras}{MNRAS}

\newcommand{\araa}{ARAA}

\newcommand{\nat}{Nature}
\newcommand{\pre}{PRE}

\renewcommand{\footnotesize}{\normalsize}

\title[Profile and timing stability of millisecond pulsars]{   Limitations in timing precision due to single-pulse shape variability in  millisecond pulsars}

\author[R.~M.~Shannon et al.]{ R.~M.~Shannon$^1$\thanks{E-mail: ryan.shannon@csiro.au}, S.~Os\l owski$^{2,3}$, S.~Dai$^{1,4}$, M.~Bailes$^5$, G. Hobbs$^1$,   R.~N.~Manchester$^1$,  \newauthor W.~van Straten$^{5}$,  C.~A.~Raithel$^{6}$, V.~Ravi$^{7,1}$, L.~Toomey$^1$, N.~D.~R. Bhat$^8$, \newauthor  S.~Burke-Spolaor$^9$, W.~A.~Coles$^{10}$,  M.~J.~Keith$^{11}$,  M.~Kerr$^{1}$, Y.~Levin$^{12}$, J.~M.~Sarkissian$^{13}$, \newauthor J.-B.~Wang$^{14}$, L.~Wen$^{15}$, X.-J.~Zhu$^{15}$  \\
$^1$ CSIRO Astronomy and Space Science, Australia Telescope National Facility, Box 76, Epping NSW 1710, Australia \\
$^2$  Max-Planck-Institut f{\"u}r Radioastronomie, Auf dem H{\"u}gel 69, D-53121 Bonn, Germany \\
$^3$  Department of Physics, Universit\"at Bielefeld, Universit\"atsstr. 25 D-33615 Bielefeld, Germany \\
$^4$  Department of Astronomy, School of Physics, Peking University, Beijing, 100871, China\\
 $^5$ Centre for Astrophysics and Supercomputing, Swinburne University of Technology, Post Office Box 218, Hawthorn, VIC 3122, Australia\\ 
$^6$  Department of Physics, Carleton College, Northfield, MN, 55057, USA \\
$^7$ School of Physics, University of Melbourne, Parkville, VIC 3010, Australia \\
$^8$  International Centre for Radio Astronomy Research, Curtin University, Bentley, WA 6102, Australia \\
$^9$  Department of Astronomy, California Institute of Technology, Pasadena, CA 91125, USA\\
$^{10}$  Department of Electrical and Computer Engineering, University of California, San Diego, La Jolla, CA 92093, USA \\
$^{11}$ Jodrell Bank Centre for Astrophysics, University of Manchester, M13 9PL, UK   \\
$^{12}$ School of Physics, Monash University, PO Box 27 Clayton, VIC 3800, Australia \\
$^{13}$  CSIRO Astronomy and Space Science, Parkes Observatory, Box 276, Parkes NSW 2870, Australia \\
$^{14}$ Xinjiang Astronomical Observatory, Chinese Academy of Sciences, 150 Science 1-Street, Urumqi, Xinjiang 830011, China\\
$^{15}$  Department of Physics, University of Western Australia, Crawley, WA 6009, Australia }

\begin{document}

\date{Accepted XX ; in original form \today}

\pagerange{1--21} \pubyear{2014}

\maketitle

\begin{abstract}

High-sensitivity radio-frequency observations  of millisecond pulsars usually show  stochastic, broadband, pulse-shape variations intrinsic to the pulsar emission process.
These variations  induce jitter noise  in  pulsar timing observations;
understanding the properties of this  noise is of particular importance for the effort to detect gravitational waves with pulsar timing arrays.
We assess the short-term  profile and timing stability of $22$ millisecond pulsars that are part of the Parkes Pulsar Timing Array sample 
by examining intra-observation arrival time variability and single-pulse phenomenology.  
In  $7$ of the $22$ pulsars, in the band centred at approximately $1400$~MHz,  we find that the brightest  observations are limited by intrinsic jitter. 
 We find consistent results,  either detections or upper limits, for jitter noise in other frequency bands.
  PSR~J1909$-$3744 shows the lowest levels of jitter noise, which we  estimate to contribute    $\sim 10$~ns root mean square error to the arrival times for  hour-duration observations.  
 Larger levels of jitter noise are found in pulsars with wider pulses and distributions of pulse intensities.
The jitter noise in PSR~J0437$-$4715 decorrelates over a bandwidth of $\sim 2$~GHz.
We show that the uncertainties associated with timing pulsar models  can be improved by including  physically motivated jitter uncertainties. 
Pulse-shape variations will limit the timing precision at future, more sensitive, telescopes; 
 it is imperative to account for this noise when designing instrumentation and timing campaigns for these facilities.     
\end{abstract}

\clearpage

\section{Introduction}

Pulsar timing measurements enable the study of myriad phenomena of fundamental astrophysical and physical interest.  These measurements, for example, have been used to characterise the orbits of binary systems,  enabling tests of general relativity  \cite[][]{2006Sci...314...97K}, constraining nuclear equations of state  \cite[][]{2010Natur.467.1081D,2013Sci...340..448A}, and detecting planetary-mass companions \cite[][]{1992Natur.355..145W}.      
By monitoring variations in pulse times of arrival (TOAs)  from an ensemble of the most stable millisecond pulsars (MSPs) that have time-of-arrival precision of $< 100$~ns, it is possible to detect the presence of nanohertz-frequency gravitational radiation \cite[][]{1979ApJ...234.1100D,1983ApJ...265L..39H}.     The ensemble  is referred to as a pulsar timing array \cite[PTA,][]{1990ApJ...361..300F}.  
Current limits on gravitational radiation have been used to constrain the growth and evolution of black holes and their host Galaxies in the low-redshift ($z \lesssim 1$) Universe \cite[][]{2013Sci...342..334S}.
In order to detect   gravitational waves it is necessary to improve the pulsar timing array datasets.  This can be accomplished by 1) observing a larger set of pulsars; 2) increasing the observing span of the observations; and 3) increasing the quality of pulsar timing array datasets \cite[][]{2012ApJ...750...89C,2013CQGra..30v4015S}.


One of the most useful diagnostics for assessing the quality of a timing model is the pulsar  timing residuals, which are the differences between the observed TOAs and a timing model \cite[e.g.,][]{2006MNRAS.372.1549E}.
It is well known that pulsar timing residuals show scatter in excess of what would be predicted by formal timing uncertainties \cite[][]{1975ApJS...29..453G}.
  This excess can be divided phenomenologically into at least two components;  a time-correlated red-noise component and a  white-noise component that is uncorrelated between observing epochs.  The red noise  can contain contributions from intrinsic spin noise \cite[][]{sc2010,2014MNRAS.437...21M}, magnetospheric torque variations \cite[][]{2010Sci...329..408L},  uncorrected dispersion variations \cite[][]{2013MNRAS.429.2161K} and multi-path propagation effects \cite[][]{cs2010} in the interstellar medium, inaccuracies in terrestrial time standards \cite[][]{2012MNRAS.427.2780H}, uncertainties in the solar system ephemeris \cite[][]{2010ApJ...720L.201C},  the presence of asteroid belts \cite[][]{2013ApJ...766....5S}, or other phenomena.

In addition to radiometer noise, white noise can originate from a number of sources.   
One of the  most significant effects is associated with the difference between the  ensemble-average pulse profile and the average of a finite number of pulses.
The difference biases the measurements of arrival times, contributing  {\em jitter} noise to the TOAs.
Single pulses for nearly every pulsar observed with high sensitivity  show variation in excess of that expected from radiometer noise.  
This includes variations in amplitude and phase that are correlated from pulse to pulse (such as the drifting sub-pulse phenomenon) and variations that are uncorrelated from pulse to pulse.    
If the jitter noise is independent from pulse to pulse (or decorrelates on a timescale shorter than the time resolution of the observations), the root mean square (rms) error $\sigma_J$  scales proportional to $\sigma_J(N_p) \propto 1/\sqrt{N_p}$, where $N_p$ is the number of pulses averaged in forming the integrated profile.     The jitter noise is characterised either in terms of its rms contribution to the residual arrival times  $\sigma_J(N_p)$ or the dimensionless jitter parameter \cite[][]{2012ApJ...761...64S}:

\be
\label{eqn:jit_param}
f_J \equiv  \frac{\sigma_J(N_p=1)}{W_{\rm eff}} 
\ee
where $W_{\rm eff}$ is the instrinsic pulse width.  \cite{2012ApJ...761...64S} suggest multiplying the effective width by the factor  $(1+m_I^2)$, where $m_I$ is the modulation index of the pulse energies.     
The motivation for including this factor is to distinguish variations in intensity from variations in shape (see \cite{2012ApJ...761...64S} and \cite{cs2010} for further discussion). 
The modulation index
can be calculated from the mean $\mu_E$ and standard deviation $\sigma_E$ from the pulse-energy distribution:
\be
m_I = \frac{\sigma_E}{\mu_E}.
\ee

We consider different measurements of the effective pulse width, including both the full widths at $50\%$  and $10\%$ of peak intensity ($W_{50}$ and $W_{10}$, respectively) and effective widths that take into account the pulse shape. 
One measure of the effective pulse width that has been suggested  \cite[][]{1983ApJS...53..169D,cs2010} is
\be
\label{eqn:effective_width_sc}
W_{\rm eff} = \frac{\Delta \phi}{ \sum_i [I(\phi_{i+1}) - I(\phi_i)]^2 },
\ee
where $\Delta \phi$ is the phase resolution of the pulse profile (measured in units of time), and the pulse profile is normalised to have a maximum intensity of unity. 
The denominator of Equation (\ref{eqn:effective_width_sc}) is proportional to the mean-squared derivative of the pulse profile and is therefore a measure of the sharpness of the pulse profile.
   Another measure of the effective pulse width that has been used  \cite[][]{2012MNRAS.420..361L} is
\be
\label{eqn:effective_width_liu}
W_{{\rm eff},L}=   \frac{\int~d\phi\, \phi^2 I(\phi)}{ \int d\phi\, I(\phi)}.
\ee
In Equations (\ref{eqn:effective_width_sc}) and (\ref{eqn:effective_width_liu}),   $I(\phi)$ is the mean pulse profile as a function of pulse phase $\phi$ (measured in units of time).


Jitter noise is well known to be present in  slower spinning  pulsars \cite[][]{1975ApJ...198..661H,1985ApJS...59..343C} and is expected to be present in all pulsar observations when the single-pulse signal to noise ratio (S/N) exceeds unity \cite[][]{2011MNRAS.418.1258O,2012ApJ...761...64S}.  Given the importance of precise timing to PTA experiments, a few recent studies have  attempted to identify  the presence of pulse jitter in  millisecond pulsars.
Using observations from  the $64$-m Parkes telescope at an observing frequency of $\sim 1400$~MHz, \cite{2011MNRAS.418.1258O}  investigated the timing precision limits in PSR~J0437$-$4715, finding that in $1$~hr of observation, shape variations limit the timing precision to approximately $30$~ns.  
Using  observations from the Parkes telescope of PSR J0437$-$4715 at an observing frequency $\sim 1400$~MHz, \cite{2012MNRAS.420..361L} found a consistent level of jitter noise and estimated the jitter parameter to be $f_J=0.04$, based on the effective width defined in Equation (\ref{eqn:effective_width_liu}).
Using observations from the $305$-m Arecibo  telescope at $\sim 1600$~MHz, \cite{2012ApJ...761...64S} connected single pulse variability in PSR~J1713$+$0747 to high precision timing observations to find that jitter contributes $\sim 20$~ns  of timing uncertainty for an hour-duration observation.   

The presence of jitter noise is connected to the stochasticity of  single pulses.    
 The single pulses of only three MSPs have hitherto been well characterised.  Not surprisingly, these are three of the brightest MSPs at decimetre wavelengths:  PSR J0437$-$4715    \cite[][]{1997ApJ...475L..33A,1998ApJ...498..365J,2014oslo}; PSR J1939$+$2134 \cite[][]{2001ApJ...546..394J,2004ApJ...602L..89J}; and PSR J1713$+$0747 \cite[][]{2012ApJ...761...64S}.
\cite{2003A&A...407..273E} detected  individual pulses  in  PSRs~J1012$+$5307, J1022$+$1001, J1713$+$0747 and J2145$-$0750; however only  $\sim 100$  pulses were detected for each pulsar and the statistics of the distribution of pulse energies were not explored.   
Additionally   \cite{2003A&A...407..273E} found evidence for quasi-periodic modulation of pulse intensities on time scales of $\sim 10$~pulse periods for PSRs J1012$+$5307 and J1518$+$4094.  These quasiperiodicites were  found not to dominate the single-pulse intensity modulation. 
In addition giant pulses, narrow pulses with energies that can be a factor of $40$ greater than the mean pulse energy have been detected from PSR~J1939$+$2134, PSR~J1824$-$2452A, and PSR~J1823$-$3021A \cite[][]{2005ApJ...625..951K}.

While single-pulse variability is a nuisance for precision timing, it can be used as a tool  to test models of the pulse emission mechanism. 
\cite{2004MNRAS.353..270C} studied the phase-resolved single-pulse properties of two slower-spinning pulsars, PSRs  B0950$+$08 and B1641$-$45, and interpreted these in the context of models of pulsar emission.
They found that over much of pulse phase, both pulsars showed log-normal energy distributions, and argued that stochastic growth, which predicts this type of distribution,   plays the central role in the production of pulsar emission, in which linear instabilities in the plasma generate the radio emission.
They contrast this theory to  non-linear growth models which  predict   power-law energy distributions.
Power-law energy distributions can also be produced from the vectorial superposition of two wave populations \cite[][]{2002PhRvE..66f6614C}.
\cite{2004MNRAS.353..270C}  also found that near the edges of the pulse profile both pulsars showed Gaussian modulation, and suggested it was caused by either refraction in the magnetosphere, the superposition of many independent (log-normal) components, or  was intrinsic to the emission mechanism. 


Previous attempts to study single pulses and giant-pulse emission from MSPs have been limited by the low expected S/N  for single pulses. 
Here we  expand on previous studies to identify pulse-shape variations and assess the levels of jitter noise in the Parkes Pulsar Timing Array MSP sample \cite[][]{2013PASA...30...17M}.   Over the duration of the project, $22$~MSPs have been regularly observed enabling us to measure or place limits on the levels of pulse jitter in these objects.  
The high cadence and long duration of the project has enabled us to select observations for which refractive and diffractive scintillation  have significantly increased the observe flux density of the pulsars, enabling us to both  detect single pulses and measure the effects of pulse jitter.
In Section \ref{sec:observations}, we present the observations. 
In Section \ref{sec:analysis}, the analysis methods that we use are discussed.
In Section \ref{sec:results}, we present results from  the PPTA pulsars. 
In Section \ref{sec:improve} we present a technique to correct TOA uncertainties for the effects of jitter noise.  We apply this technique to a multi-year observations of PSR~J0437$-$4715. 
We discuss and summarise our findings in Section \ref{sec:discuss}.  

\section{Observations} \label{sec:observations}

For our analysis, we selected  observations from the PPTA project, which includes  observations   of  $22$  MSPs south of a declination of $\approx +24^\circ$, the northern declination limit of the Parkes antenna.  
The pulsars are observed regularly, with an approximate observing cadence of  three weeks, in three bands centred close to $730$ MHz, $1400$~MHz, and $3100$~MHz, using the dual-band 10cm/50cm receiver and the central beam of the 20cm multibeam receiver.
   In each of the bands, the observing bandwidth is $64$~MHz, $256$~MHz, and $1024$~MHz, respectively.   
   While the 20cm system is typically the most sensitive to single pulses and pulse jitter, we also analysed observations  obtained with the 10cm/50cm system to search for, or place limits on, these effects.

Most  of the pulsars in the sample show large flux density variability at the PPTA observing frequencies due to diffractive and refractive interstellar scintillation \cite[][]{1990ARA&A..28..561R}.
Diffractive interstellar scintillation  causes pulsar radiation to show time and frequency variability in which the dynamic spectrum is broken up into scintles.  
Individual scintles show exponential distribution of intensity statistics and therefore have a long tail of rare but high intensities. 
In observing bands populated by few scintles, flux measurements show exponential or nearly-exponential distribution in intensity. 
 Refractive scintillation causes magnification (or de-magnification) of this pattern as detected at Earth, causing further variation in  intensity.  
We find that some  of the pulsars in the PPTA sample show measured intensities a factor of 20  greater than  the mean. For these observations the    Parkes observations have an S/N  representative (or in excess) of the average observations  of larger-aperture telescopes such as the Green Bank Telescope and the  expected observations from  the MeerKAT telescope.

\subsection{Fold-mode observations } \label{sec:foldmode}

In standard pulsar timing observations, spectra are formed and folded at the pulse period of the pulsar, as predicted by its ephemeris. 
For our observations, spectra were formed using both digital polyphase filterbank spectrometers, (PDFB3 and PDFB4);  and coherent dedispersion machines (CPU-driven APSR and  GPU-driven CASPSR).  
Observations of this type form the basis of the PPTA dataset and comprise one component of the data analysed here.
Individual subintegrations were of $8$~s or $32$~s duration for CASPSR, and $60$~s duration for the other backends.
For further details see \cite{2013PASA...30...17M} and references therein.

Data calibration  was conducted using standard data reduction tools \cite[][]{2004PASA...21..302H}.
To excise radio-frequency interference,  we median-filtered each sub-integration in the frequency domain. 
The polarisation was calibrated by correcting for differential gain and phase between the receptors through measurements of a noise diode injected at an angle of $45^\circ$ from  the linear receptors. 
  In  some observations with the 20cm~system, we corrected for cross  coupling between the feeds through a model derived from an observation of PSR~J0437$-$4715 that covered a wide range of parallactic angles \cite[][]{2004ApJS..152..129V}. 
 However,  we find that our results were independent of this cross-coupling calibration; this is because the effects of polarisation  are small compared the levels of jitter in our short ($\lesssim 1$~hr) observations  that cover   a small range in parallactic angle.
The observations were then flux calibrated using observations of the radio galaxy Hydra~A, which is assumed to have a constant flux and spectral index \cite[][]{1968ARA&A...6..321S}
 
We calculated TOAs  by cross correlating frequency-averaged observations with a template  in the Fourier domain \cite[][]{1992RSPTA.341..117T},  which is presently  the most common algorithm used for measuring arrival times.
This algorithm assumes that the only source of noise in the measurement is white noise.  
The formal TOA uncertainties, $\Delta_F$, are based on this assumption and therefore underestimate the true TOA uncertainty \cite[][]{2011MNRAS.418.1258O}.

\subsection{Baseband  observations}

We recorded raw-voltage  (baseband) data for short intervals when pulsars were identified to be in particularly bright scintillation states. 
  These intervals were identified in real time when  the single-pulse S/N (measured by extrapolating from the fold-mode observations)  significantly exceeded unity.
 Baseband data were recorded with the CASPSR instrument, which is capable of simultaneous real-time coherent dedispersion and baseband recording.
  Full-Stokes single-pulse profiles were created by  coherently desdispersing  the baseband data off-line  \cite[][]{2011PASA...28....1V} and calibrating for differential gain and phase of the feeds, and correcting for their cross coupling where appropriate \cite[][]{2013ApJS..204...13V}.  These observations  were not flux-calibrated.

In Table \ref{tab:single_pulse_obs}, we summarise the seven single-pulse datasets used in this analysis. 
For all of the pulsars, between 40,000 and 300,000 pulses were observed. 
All of the single-pulse observations were obtained with the 20cm system.

\begin{table*} 
\caption{ \label{tab:single_pulse_obs}  Single-pulse observations. } 
\begin{center}
\begin{tabular}{lrrcccc}
\hline \hline
PSR & $P$ & $S_{1400} $ &  MJD & $N_p$ & $\langle {\rm S/N} \rangle$ & S/N$_{\rm max}$ \\ 
 & (ms) & (mJy) & & \\
\hline
J0437$-$4715 &  $5.76$   & $149$    &  $56446$    & $1.0 \times 10^5$  & $16$ & $89$\\
J1022$+$1001&  $16.45$ &    $6$     &   $56304$   & $3.8 \times 10^4$  & $1.8$ & $9.9$ \\
J1603$-$7202 &  $14.84$ &     $3$    &  $56409$    & $4.2 \times 10^4$  & $ 1.6$ & $11$  \\
J1713$+$0747 & $ 4.57$  &   $10$    & $56447$     &  $1.1 \times 10^5 $ & $1.9$ & $7.5$ \\
J1744$-$1134 &  $  4.07$ &    $3$     &   $56514$   &   $6.1 \times 10^4$  & $3.1$ & $11$ \\
J1909$-$3744 & $2.95$   &   $2$       &    $56310$  &  $3.9 \times 10^5$   & $2.2$ & $11$\\
J2145$-$0750 & $16.05$ &   $9$      &  $56206$     &  $ 4.3 \times 10^4$  &  $5.5$ & $22$  \\
\hline
\end{tabular}
\end{center}
Notes:    For each pulsar, we list the period $P$ of the pulsar, the flux density $S_{1400}$ at a frequency of $1400$~MHz, the MJD of the observation, the number of pulses obtained $N_p$,  the average  S/N ($\langle {\rm S/N} \rangle$) and the maximum S/N observed for a single-pulse S/N$_{\rm max}$.  The flux density measurements are from \cite[][]{2013PASA...30...17M}. 
 \end{table*} 

\section{Analysis methods}  \label{sec:analysis}
 
\subsection{Timing analysis}\label{sec:timing}

Using the techniques described above,  we derived TOAs from pulse profiles formed from $N_p=1$ pulse to $N_p\sim10^5$  pulses.
For each pulsar, these TOAs were fitted to  long-term timing models  derived from PPTA observations \cite[][]{2013PASA...30...17M}.

In some hour-duration fold-mode observations, we identified secular trends in arrival times.   
We attribute these trends to  pulse-shape distortions caused by diffractive interstellar scintillation  and  intrinsic pulse profile evolution \cite[][]{2014arXiv1402.1672P}.
The diffractive scintillation pattern causes variable weighting of the pulse profile  with frequency.  
If the pulse profile varies with frequency (as is common) the frequency averaged profile will change shape.  
   We find that these trends  could be adequately removed by re-fitting the timing model for pulsar spin frequency and frequency derivative.   We defer discussion of the origin of these trends and methods for mitigation to future work.

To  determine the level of jitter noise, we compared the measured rms of the residuals to levels expected from simulations of ideal datasets.  
 In these simulated datasets, we formed pulse profiles from the template and white noise such that the  S/N of each simulated sub-integration matched the observed S/N.  

We then define the rms uncertainty associated with jitter of   $N_p$ pulses averaged together, $\sigma_J(N_p)$  to be the quadrature difference between the rms of the observed and simulated datasets: 
\be
\label{eqn:jit_equation}
\sigma^2_J(N_p)  = \sigma_{\rm obs}^2(N_p)  - \sigma_{\rm sim}^2(N_p).
\ee
We assume here  that all of the excess error in the arrival time measurements can be attributed to pulse jitter.  

As discussed in  \cite{2012ApJ...761...64S}, there are other perturbations to pulse arrival times  that manifest on short (millisecond to hour) time scales;  however very few effects can cause short time-scale distortions that depend at most weakly with frequency with the same strength in many  backend instruments, at different telescopes, and at different observing frequencies, as is presented below.
Distortions in pulse profiles caused by polarisation calibration are likely to vary slowly with parallactic angle as receiver feeds rotate with respect to the pulsar \cite[][]{1984ApJS...55..247S}.  
Distortions introduced at the telescope will be observatory dependent,  backend dependent, or both.
Similarly radio frequency interference will depend on the observing band and telescope site.
By linking shape variations and timing variations on the shortest timescales to timing variations on longer timescales via Equation (\ref{eqn:jit_equation})   we estimate the contribution of jitter to TOA uncertainties.


We compared this to estimations for the level of jitter noise  presented in \cite{2012MNRAS.420..361L}.
Instead of using simulations to infer the levels of jitter $\sigma_J$, \citet{2012MNRAS.420..361L}   modify the TOA uncertainties $\Delta_{{\rm TOA}, i}$, from the formal values $\Delta_F$ for $i=1,N_{\rm obs}$  using 
\be
\Delta_{{\rm TOA}, i}^2 = \Delta_{F}^2  + \sigma^\prime_J(N)^2,
\ee
and setting  $\sigma^\prime_J(N)$ to be the value at which the  reduced $\chi^2$ of the best-fitting model  was unity. 
The jitter parameter is then calculated using Equation (\ref{eqn:jit_param}). 
We obtained consistent results for  $\sigma_J(N_p)$ using this method.

When we have single-pulse datasets, formal uncertainties
on $\sigma_J(1)$ are small because there are many independent estimates  of $\sigma_J(N_p)$. 
The uncertainty  in $\sigma_J(1)$ cannot be derived from fitting the relationship  $\sigma_J(N_p) = \sigma_J(1)/\sqrt{N_p}$ to a single dataset and multiple $N_p$  because $\sigma_J(N_p)$ are dependent and therefore their uncertainties are correlated.   Instead it has to be derived from a single measurement of $\sigma_J(N_p)$ if only one dataset is used.  Alternatively  it can be derived from multiple $\sigma_J(N_p)$ if independent datasets are used.


\subsection{Pulse-shape analysis}\label{sec:sp}

 The most direct way to link timing variations to shape variations  is to analyse the properties of single pulses or sub-integrations comprising as few pulses as feasible.
 In many observations, it was not possible to characterise every single pulse because of large variations in pulse amplitudes.
The instantaneous S/N  was typically  $\sim 1$ to $5$; the pulsars show long positive tails of pulse energies in which the  brightest pulses exceed the average pulse energy  by a factor of  $5$.

While there are  many  established tools for analysing the properties of single pulses,
central to our analysis is the measurement of the energy contained in a pulse or a sub-component of a pulse.
We define the energy of the pulse (or sub-component) to be its integrated flux density.
These included larger windows around the entire main pulse feature, subcomponents of the main pulse feature, precursor components, and interpulses.
Regions around main pulses, precursors, or interpulses were set to contain more than $90\%$ of the pulse energy.  Windows around other components were set to be centred on the component.  
If more than one sub-component was measured for a pulsar, where reasonable, we chose windows of the same size to most directly compare the statistics of the individual components.  
For each of the regions,  we also defined an off-pulse window that was used as a control sample to assess the statistics of the noise in our measurements.
   The off--pulse  windows were chosen to have the same width as the components of interest, and enabled us to empirically derive the noise statistics of the pulses.     We normalised the measured energies by subtracting the off-pulse mean and then dividing by the off-pulse standard deviation. In our plots below, we therefore measure pulse energy in units of the S/N.  
Single pulses most severely affected by RFI  showed anomalously high pulse energy and, by inspection, were removed from our sample.   These were identified as containing non-dispersed impulsive signals that affected a larger range of pulse phase than the pulsar emission. In total, for each pulsar  fewer than $10$ pulses were removed, representing $\ll 0.1\%$ of the total pulse sample.

In order to assess the intrinsic energy distribution, it is necessary to deconvolve the effects of radiometer noise.
Because the measured pulse energy is  the sum of noise and the signal, the probability density function (PDF) for the measured energy  $\rho_E$ is the convolution of the PDFs of the noise ($\rho_N$) and intrinsic energy distribution ($\rho_I$):
\be
\label{eqn:convolution}
\rho_E(E)  = \int dE^\prime \rho_N(E^\prime) \rho_I(E-E^\prime).
\ee

We find, unsurprisingly, that the off-pulse distribution   $\rho_N(E)$ was very well modelled by  a normal Gaussian.  

We consider different models for the pulse-energy histogram based on generalised log-normal distributions and generalised Gaussian distributions.
In most cases, we find that the pulse-energy histogram could be well modelled using a generalised  log-normal distribution:
\be
\label{eqn:lognormal}
\rho_I(E) = A  \exp\left( - \left|\frac{\ln E - \ln \mu_E }{\ln \sigma_E} \right|^\alpha \right),
\ee
where $A$ is a  constant that normalises the integral of the PDF to unity,  and $\ln \mu_E$,  $\ln \sigma_E$, and $\alpha$ parametrize the distribution.    For $\alpha =2$, Equation (\ref{eqn:lognormal}) is a log-normal distribution.

As discussed below, one pulsar, PSR~J1909$-$3744, shows a pulse-energy distribution with a shape that is better matched by a generalised Gaussian distribution.
In this case, the pulse-energy distribution  is modelled to be
\be
\label{eqn:gaussian}
\rho_L(E)  = A \exp\left( - \left| \frac{E - \mu_E}{\sigma_E}\right|^\alpha  \right), 
\ee
where $A$ again is the normalising constant, and $\mu_E$, $\sigma_E$, and $\alpha$ parametrize the distribution.

In order to find the best-fitting parameter values, we used a Metropolis-Hastings algorithm  \cite[][]{2005blda.book.....G} to sample the parameter space using the likelihood function for the observed pulse energy given a set of model parameters.

In each bin ($i=1,N$), centred at energy $E_i$ and of width $\Delta E$, the number of pulses  is modelled to have  multinomial probability; therefore the logarithm of the likelihood function is 
 \begin{equation}
\label{eqn:loglik}
 \log L = \log N!   + \sum_i^N   n_{i}  \log p_i   -  \log n_{i}!  
 \end{equation}
where $n_{i}$ is number of  pulses detected in  bin $i$,  and  $p_i$ is the  probability of finding a pulse in bin $i$.

Markov chains were used to sample the parameters $\bm{P} \equiv (\mu_e, \sigma_E, \alpha)$ for both the generalised log-normal and generalsed Gaussian random variables.
The Markov chain was computed using the standard procedure.  
At each step in the chain the likelihood  $L_k$ was calculated using Equation (\ref{eqn:loglik}). 
A provisional set of parameters $P^\prime_k$ were generated that were a perturbation on the previous parameters: 
\be
\bm{P}^{\prime}_k = \bm{P}_k + \bm{\Delta P}.
\ee
The perturbation $\bm{\Delta P}$ was generated from a multi-dimensional Gaussian distribution with zero mean and variances set so that the acceptance rate (described below) was approximately $0.2$ to $0.3$.  
The likelihood function  $L^{\prime}_k$ was calculated for these provisional values.  
If $L^{\prime}_k > L_k$ the step was accepted.
If $L^{\prime}_k < L_k$ the step was accepted with probability $L_k/L^{\prime}_k$.     

After a burn-in period that is used to find the global maximum of the likelihood function,  the Markov chain models the PDF of the parameters.
 For the best-fitting values we therefore take  the mean of the Markov chain and for the parameter uncertainties we take the standard deviation of the Markov chain. 
We find that the resulting best-fitting distributions  well modelled  the pulse-energy distributions. 

We used a  $\chi^2$ test statistic to assess the goodness of fit.  For our histograms the test statistic is
\be
\chi^2 = \sum_i^N  \frac{ ( n - N p_i)^2}{N p_i (1-p_i)}.
\ee
The denominator is the expected variance for bin $i$. 
The null hypothesis is that the data match the model.  
Under the assumption that the central limit theorem applies, the test statistic follows a $\chi^2$ distribution with $N_{\rm dof} = N-N_{\rm fit}$ degrees of freedom, where $N_{\rm fit}=3$ is the number of model parameters.
If the fit is good, $\chi^2/N_{\rm dof} \approx 1$.

\section{Results} \label{sec:results}

\subsection{PSR~J1713$+$0747}

\begin{figure}
\begin{center}
\includegraphics[scale=0.5]{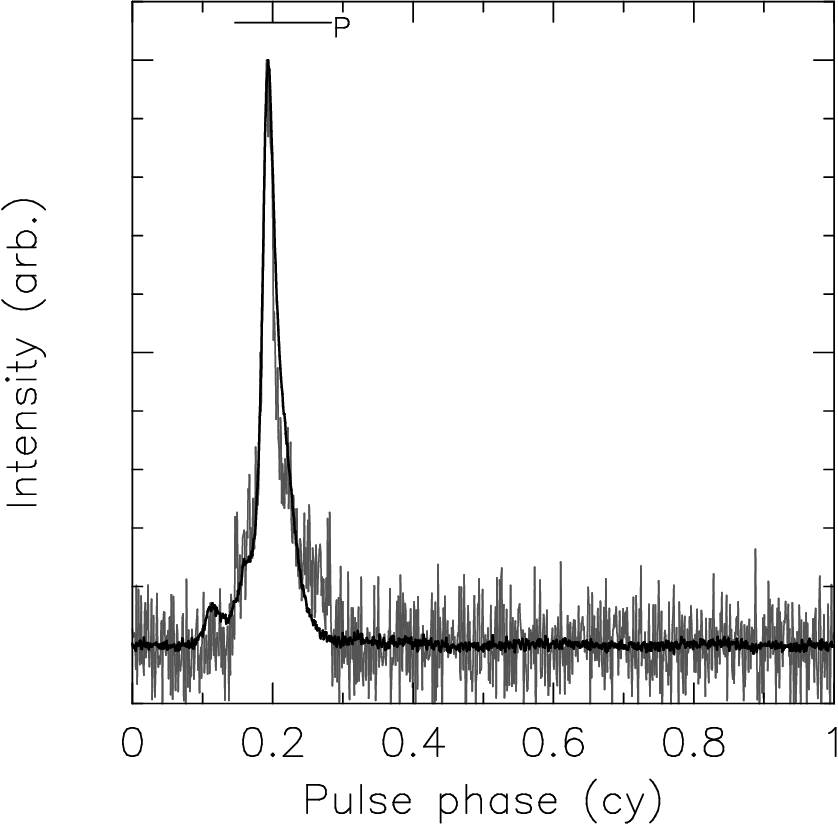}  \\
\caption{\label{fig:profiles_1713} Pulse profiles for PSR~J1713$+$0747.  The thick line shows the   average pulse profile for PSR~J1713$+$0747, derived from averaging our single-pulse observations.   The thinner grey line shows the pulse profile formed from $100$ most energetic single pulses for PSR~J1713$+$0747.   The profiles have been normalised to have the same peak flux density.  The horizontal line labelled P shows the pulse window used to measure pulse energy (see Figure \ref{fig:energy_hist_J1713}).       }
\end{center}
\end{figure}

\begin{figure}
\begin{center} 
\begin{tabular}{c}
\includegraphics[scale=0.5]{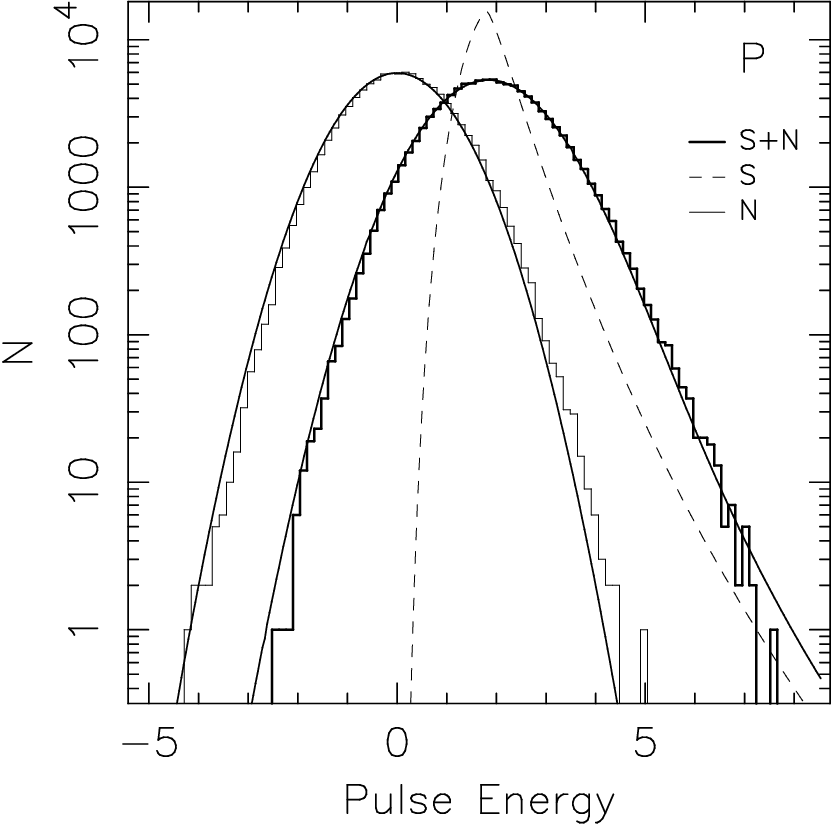} 
\end{tabular}
\caption{  \label{fig:energy_hist_J1713}  \footnotesize    Pulse-energy histograms for PSR~J1713$+$0747.  The thick solid histogram shows the pulse-energy histogram in the on pulse window.  The thick solid line (labelled S$+$N) is the  best-fitting model to the distribution.   The on pulse window is labelled P in Figure \ref{fig:profiles_1713}.   The thin solid  histogram and line (labelled N) show, respectively,  the histogram for the off-pulse window and the predicted normal Gaussian distribution.  The units of pulse energy have been scaled to the rms of the off-pulse window.     The thin dashed line (labelled S) shows the intrinsic pulse-energy histogram, deconvolved  using Equation (\ref{eqn:convolution}).      }  
\end{center}
\end{figure}

\begin{figure}
\begin{center} 
\includegraphics[scale=0.5]{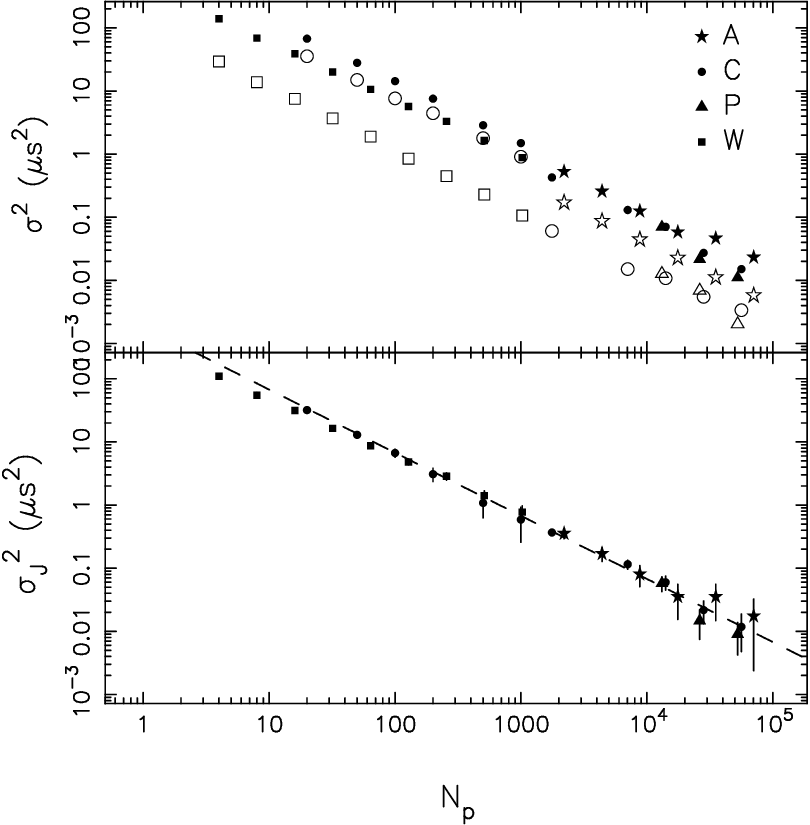} 
\caption{  \label{fig:emission_noise}  \footnotesize      Estimates of jitter noise in PSR~J1713$+$0747.   Upper panel: Variance of residual time series versus number of pulses averaged $N_p$ for observations (filled symbols) and simulated datasets (open symbols) from observations with the Parkes Telescope and the Arecibo Telescope.  Because the data were obtained with telescopes with different sensitivities, the  observed and simulated time series contain different
 levels of white noise.  Lower panel:  $\sigma_J(N_p)$, the quadrature difference between the observed and simulated datasets.  The dashed line is the best fitting model for the jitter noise scaling  $\propto N_p^{-1/2}$.   Symbols: squares - Arecibo/WAPP Observations (labelled W, autcorrelation spectrometer) observing at $1600$~MHz;  stars - Arecibo/ASP  (labelled A, a coherent dedispersion machine, using single channel of $4$~MHz of bandwidth) observing at $1400$~MHz ;   circles --  Parkes/CASPSR observing at $1400$~MHz (labelled C); triangles -- Parkes/PDFB4 observing at $1400$~MHz (labelled P)}  
\end{center} 
\end{figure}

\begin{figure}
\begin{center} 
\includegraphics[scale=0.6]{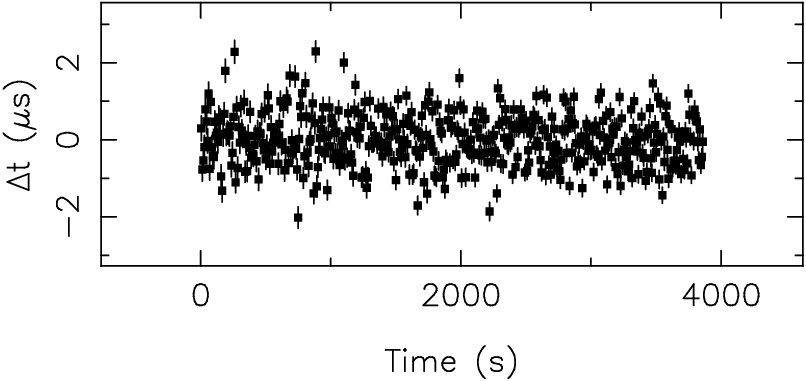}  \\
\includegraphics[scale=0.6]{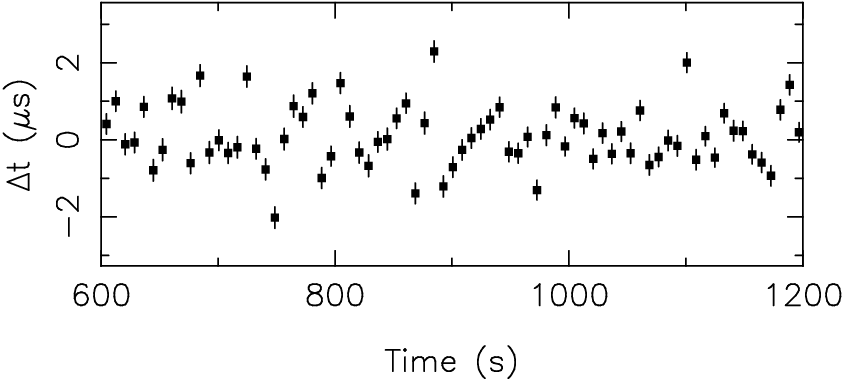}
\caption{  \footnotesize \label{fig:resid_1713}   Top panel:  Residual arrival times for $3840$~s observation of PSR~J1713$+$0747, derived from CASPSR observations with $8$~s subintegrations.  Bottom panel:  Sub-interval highlighting drift in arrival times over $\sim 48$~s at a time of $900$~s.    }  
\end{center} 
\end{figure}

In an important test, our pulse-shape analysis   of  PSR~J1713$+$0747 is consistent with a  previous analysis  presented in  \cite{2012ApJ...761...64S}. 

At $1400$~MHz, the pulse profile of PSR~J1713$+$0747,  displayed in  Figure \ref{fig:profiles_1713}, is dominated by a $100~\mu$s wide component flanked by broader  emission.  
 While the brightest single pulse has only an S/N of $\approx 10$, 
the presence of the bright pulses is sufficient to distort averaged pulse shapes and induce excess scatter in  the residual arrival times. 
The average pulse profile of the $100$ brightest pulses is also displayed in Figure \ref{fig:profiles_1713}.
Compared to the average of all of the pulses, the profile is narrower with the peak of the profile located towards the  leading edge of the average profile.
The pulse width inferred from the average of all of the pulses is $110~\mu$s, whereas the $50\%$ pulse width from the brightest pulses is  $\approx 92~\mu$s. 
When cross-correlating the brightest pulses with the average profile, we find that the bright profile is shifted early by  $ 8.1 \pm 0.1~\mu$s.  These results are consistent with  observation of the correlation between S/N and early arrival time found by \cite{2012ApJ...761...64S}.

The energy distribution of single pulses, displayed in Figure \ref{fig:energy_hist_J1713},   shows that the pulse energies have approximately a log-normal energy distribution.
 The best-fitting model parameters for the pulse-energy distributions, for this and other pulsars,  are displayed in Table \ref{tab:energy_models}.  In particular we find that $\alpha \approx  1.4$, rather than $2$ expected for a log-normal distribution.  
This energy distribution is in general agreement with observations made in the same frequency band by \cite{2012ApJ...761...64S}.  
Based on the model energy distribution we find that the modulation index  is $m_I \approx 0.3$ averaged over a window encompassing most of the pulse energy. 
While this is a borderline value for Gaussian intensity modulation \cite[][]{2004ApJ...606.1154M}, the pulse-energy statistics clearly depart from Gaussianity.

In Figure \ref{fig:emission_noise}, we compare estimates of the levels of jitter $\sigma_J(N_p)$ from these Parkes observations to the previous Arecibo observations  \cite[][]{2012ApJ...761...64S}.  
While the observations show different levels of total timing error, displayed in the  upper panel of Figure \ref{fig:emission_noise},     this is entirely due to the different sensitivity of the observing systems and the scintillation state of the pulsar at the epochs of observation. 
 After subtracting the contribution associated with radiometer noise, the excess noise can be modelled using a single power-law,  shown in the bottom panel of Figure \ref{fig:emission_noise}.   
We emphasise again that the previous study utilised observations from the Arecibo telescope and two backends with markedly different architectures than the ones used here:  including an autocorrelation spectrometer and  a CPU-based coherent dedispersion machine with lower frequency resolution. 
 Based on both analyses we expect jitter to contribute $\sim 25$~ns of rms uncertainty to an hour-long observation of PSR J1713$+$0747.

We were able to detect jitter noise in PSR~J1713$+$0747 at $3100$~MHz using only fold-mode observations.  We find that the rms contribution of jitter noise was similar at $1400$~MHz and $3100$~MHz.  At these frequencies the average pulse profiles have similar widths.  
We did not detect jitter noise in $730$~MHz  observations.  Our limit level of jitter noise is larger, and hence consistent, with the measured level in the higher-frequency observations

We also find evidence for time-correlated structure in the residuals.
In  the brightest CASPSR fold-mode observation at $1400$~MHz,  the residuals occasionally show monotonic drifts across $\sim 1~\mu$s over $\sim 48$~s.
In Figure \ref{fig:resid_1713} we show  the residual arrival times  for both the entire $\sim 3800$~s  observation  and a subsection showing an apparent drift.
The drifts are correlated between both the top half of the sub-band, indicating that pulse profile evolution modulated by the dynamic spectrum is not causing this effect.
One other pulsar, PSR~J1909$-$3744 shows structure in the residuals, with a different magnitude on a shorter time-scale.  This is discussed further below. 
Other pulsars,  with comparable or better timing precision, such as PSR~J0437$-$4715 discussed below, do not show a drift like this, suggesting that the effect is not associated with the backend instrumentation or data analysis. 
We were unable to detect this effect in other backend instrumentation because of insufficient time resolution. 
This emission could possibly be associated with the drifting sub-pulse phenomenon,  observed in many slower pulsars, in which bright emission gradually moves through the pulse profile, with an inferred drift rate of $\ll 0.1$ cycles per pulse period.  It could also be aliased from a much higher drift rate.
If unaliased, the inferred drift rate is lower by at least six orders of magnitude  than that observed in slower pulsars.
This drifting is subdominant to the random white-noise component to pulse jitter.

\subsection{PSR~J0437$-$4715}

\begin{figure}
\begin{center}
\includegraphics[scale=0.5]{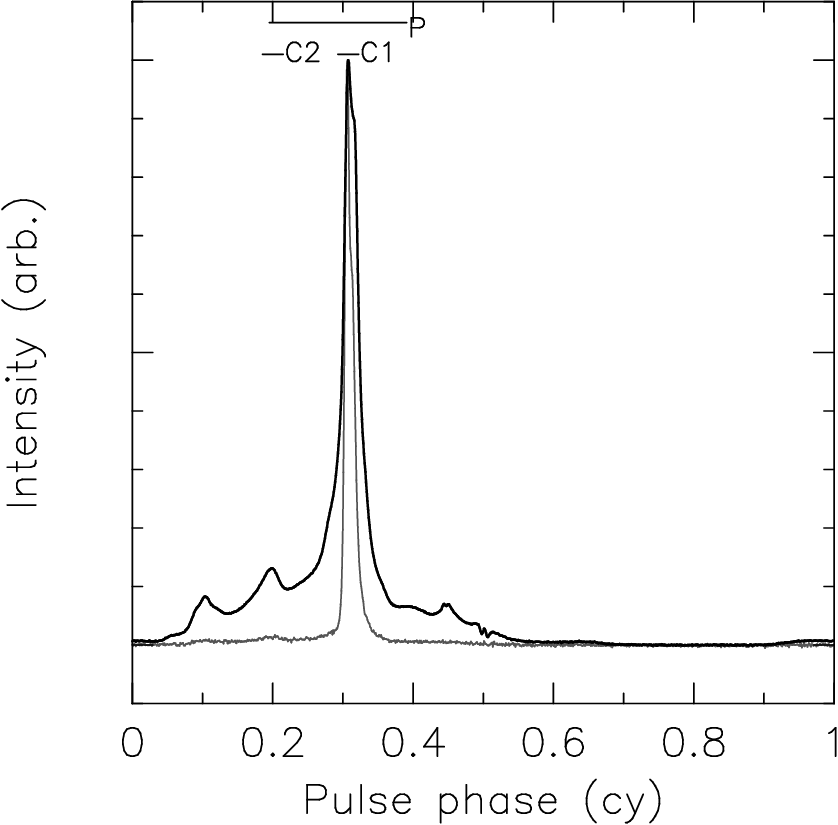}  \\
\caption{\label{fig:profiles_0437}   Pulse profiles for PSR~J0437$-$4715. The thick line shows an average pulse profile for PSR J0437$-$4715 formed from all of our single-pulse observation.   The thiner grey line shows the profile formed from $100$ most energetic  pulses.    The profiles have been normalised to have the same peak flux density.}
\end{center}
\end{figure}

\begin{figure}
\begin{center} 
\includegraphics[scale=0.5]{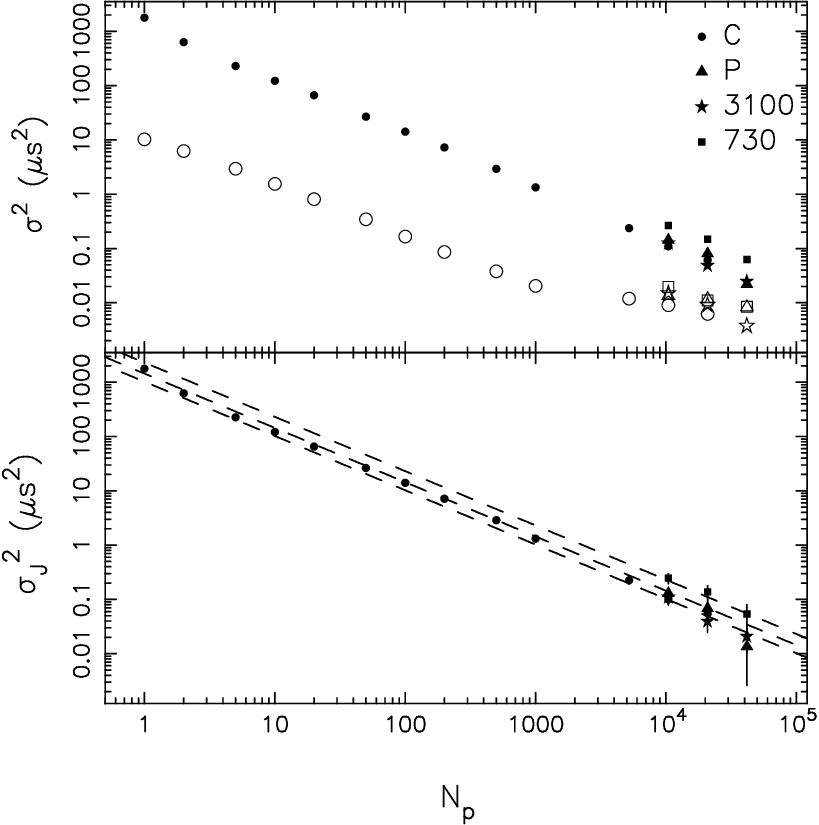} 
\caption{  \label{fig:emission_noise_0437}  \footnotesize      Estimates for levels of  jitter noise in PSR~J0437$-$4715.   Upper panel: Variance of residual time series for  observed datasets (filled symbols) and simulated datasets (open symbols) from observations with the Parkes Telescope.  
  Lower panel:  The quadrature difference between the observed and simulated datasets,  $\sigma_J(N_p)$.   Symbols: squares - 50cm/PDFB3 observations  at $730$~MHz (labelled 730);  stars - 10cm/PDFB4 observations  observing at $3100$~MHz (labelled 3100);   circles --  Parkes/CASPSR observing at $1400$~MHz (labelled C); triangles -- Parkes/PDFB4 observing at $1400$~MHz (labelled P).  The top dashed line shows the fitted jitter model for the 50cm data.   The middle dashed line shows the fitted jitter noise to the 20cm data.  The bottom dashed line shows the fitter jitter model for the 10 cm observations  }  
\end{center} 
\end{figure}


\begin{figure}
\begin{center} 
\begin{tabular}{ccc}
\includegraphics[scale=0.4]{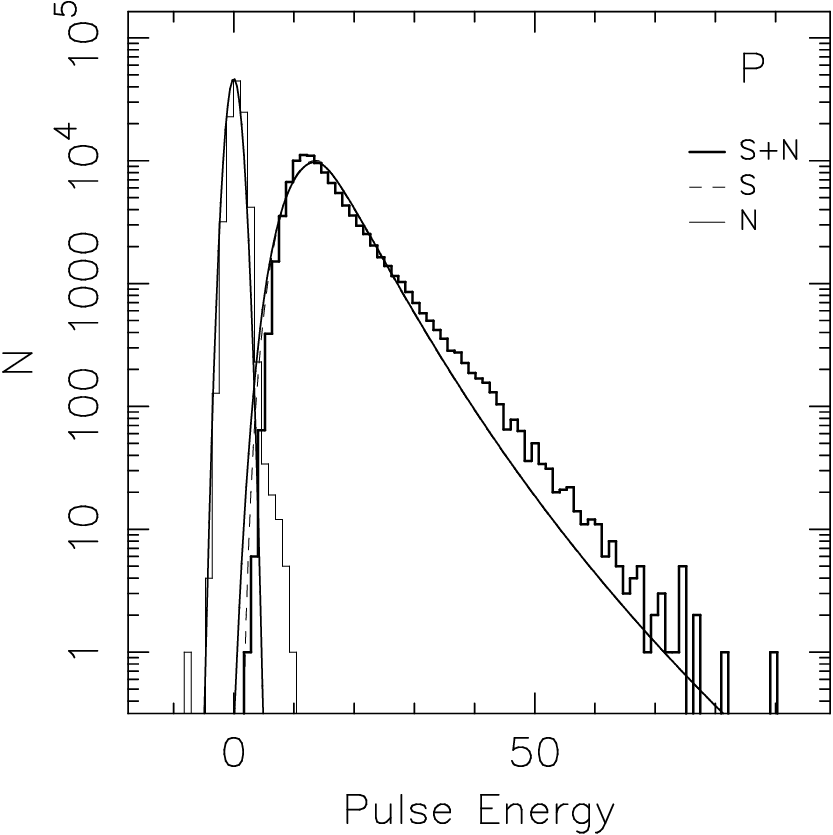} \\
\vspace{2mm}
\includegraphics[scale=0.4]{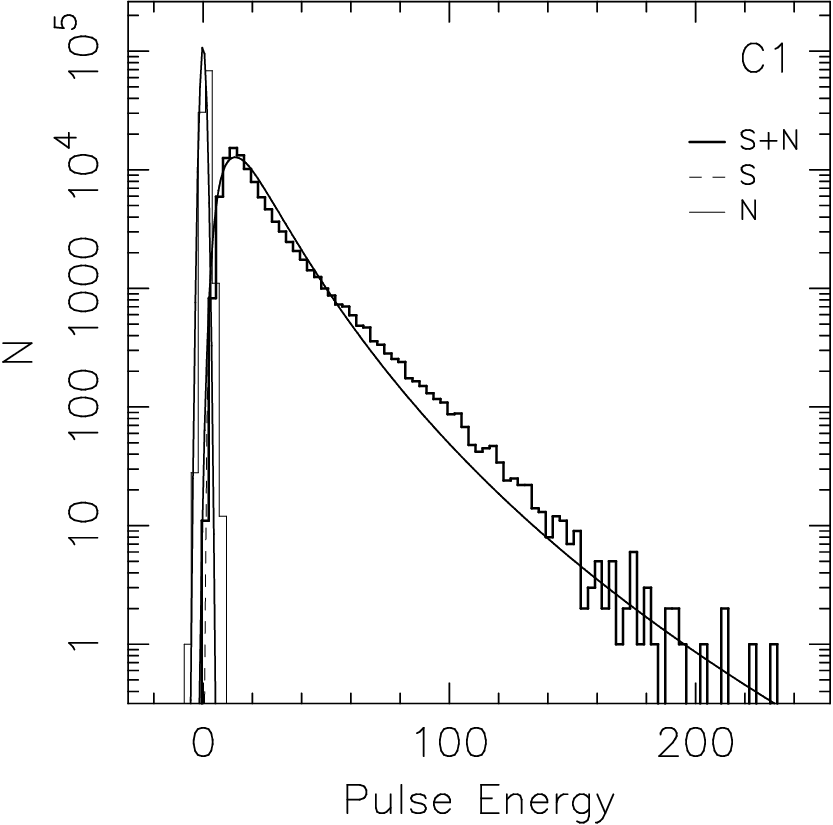}  \\
\includegraphics[scale=0.4]{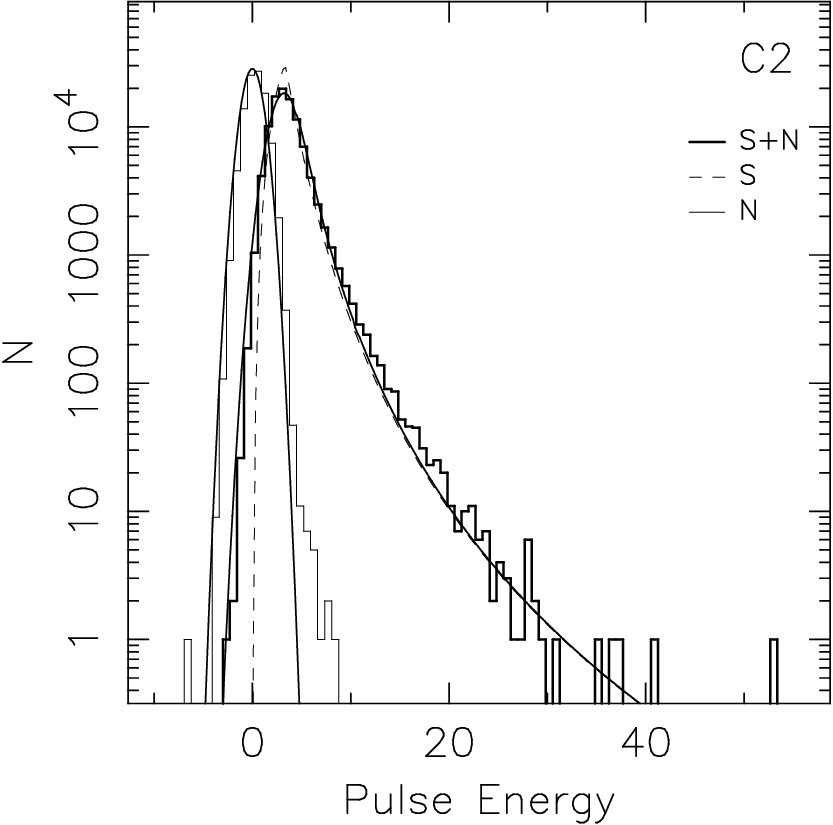} \\
\end{tabular}
\caption{  \label{fig:energy_hist_J0437}  \footnotesize    Pulse-energy histograms for PSR J0437$-$4715.  The uppermost panel shows the histogram for a window containing the brightest part of the  pulse profile (labelled P in the upper panel of Figure \ref{fig:profiles_0437}), the middle panel shows the distribution for a window centred on  component C1, and the lowermost panel shows the distribution for a window centred on  component C2.         The labels are the same as in Figure \ref{fig:energy_hist_J1713}.   }  
\end{center}
\end{figure}

\begin{figure}
\begin{center} 
\includegraphics[scale=0.4]{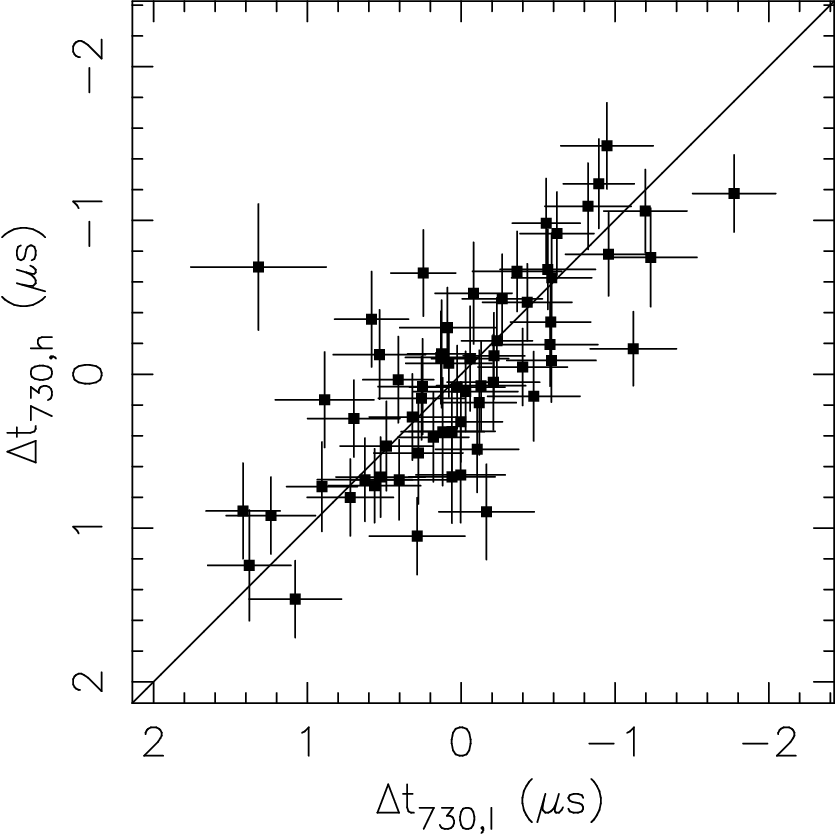} 
\vspace{4mm}
\includegraphics[scale=0.4]{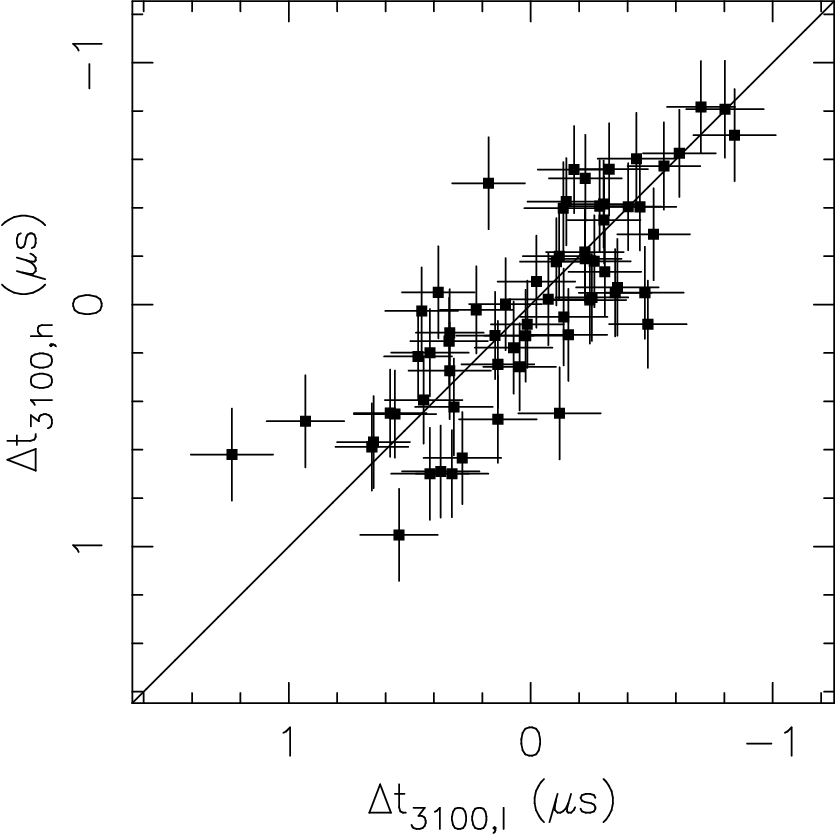}  \\
\vspace{4mm}
\includegraphics[scale=0.4]{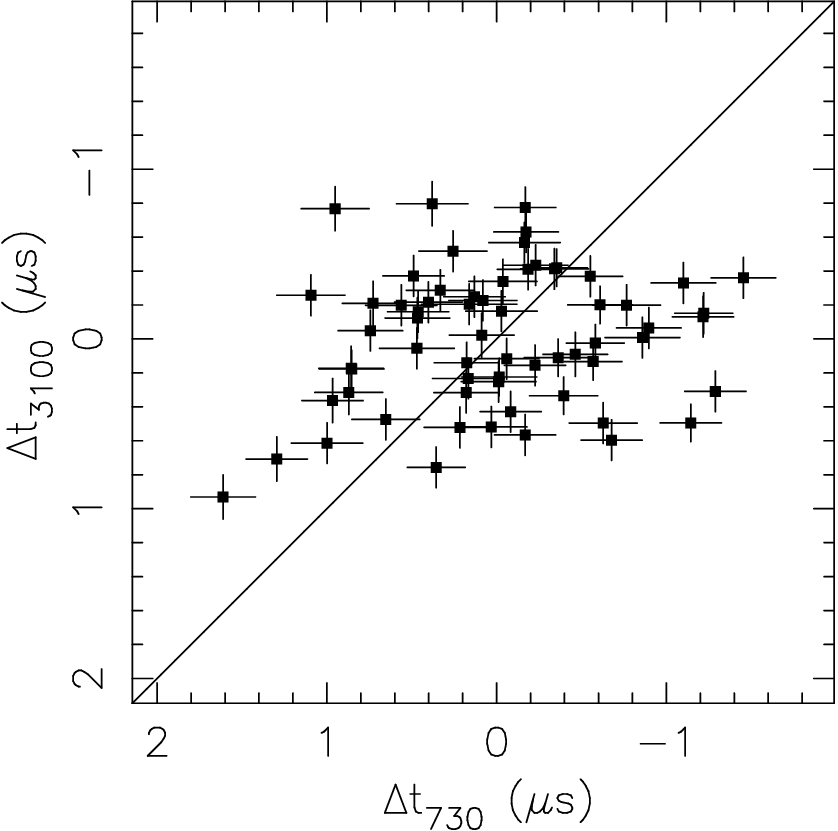} \\
\caption{  \label{fig:resid_corr_J0437}  \footnotesize    Correlation of residual TOAs for PSR~J0437$-$4715.  In the uppermost panel, we show the correlation between residuals  formed from the  lower half   $\Delta t_{730,l}$ and  upper half  $\Delta t_{730,h}$ of the $730$~MHz  (50cm) band.      In the middle panel, we show the correlation between residuals formed from the  lower half   $\Delta t_{3100,l}$ and  upper half  $\Delta t_{3100,h}$ of the $3100$~MHz (10cm) band.  
In the bottom panel, we plot residuals  formed at $730$~MHz  ($\Delta t_{730}$) and $3100$~MHz ($\Delta t_{3100}$) observations.   
  In all of the panels,  the solid lines denote unit correlation.   }  
\end{center}
\end{figure}

At decimetre wavelengths,  PSR~J0437$-$4715 is the brightest known MSP, with a phase-averaged  flux density of  $150$~mJy at $1400$~MHz.  
Because of its high flux density, pulse-shape variations cause  timing uncertainty  that is at least a  factor  $8$ greater than that  expected from  radiometer noise \cite[][]{2011MNRAS.418.1258O}.
 Its single pulses have been  widely studied (see references above). 

At $1400$~MHz, the pulsar has detectable emission  over  more than $85$\% of pulse phase.  The average pulse profile, displayed in Figure \ref{fig:profiles_0437},  shows many components but is dominated by a central peak.     In  Figure \ref{fig:profiles_0437}, we also show a pulse profile formed by averaging the $100$ most energetic pulses.  The   $50\%$ pulse width of the average of the brightest pulses is $80~\mu$s, which is significantly narrower than the $140~\mu$s width of all of the pulses.  These are consistent with observations presented in \cite{2011MNRAS.418.1258O}.

For PSR~J0437$-$4715, we analysed windows encompassing most of the pulse energy centred on  the main peak (labelled P in Figure \ref{fig:profiles_0437}) and narrower windows centred on the main peak and two of the leading sub-components.  
The pulse-energy distributions for  emission in windows P, C1, and C2  are displayed in Figure \ref{fig:energy_hist_J0437}. 
We find that for all three components, the pulse shows approximately log-normal pulse-energy distribution, though the best-fitting models, listed in Table \ref{tab:energy_models}, are the worst matching of all of the pulsars in our sample.  For  windows P  and C1, we find an excess of intermediate strength pulses.  Based on the calculations of models of pulse energy we find that the modulation index  $m_I$ for the components varies from $0.3$ to $0.7$ with the  modulation index over the widest window being $0.4$. 
We searched for,  but did not find evidence of, correlations in energy between the different windows.


Like PSR~J1713$+$0747, we determined the level of jitter noise $\sigma_J(N_p)$ for PSR~J0437$-$4715.
The results  of this analysis are presented in Figure \ref{fig:emission_noise_0437}.
We find that both the total timing error  and $\sigma_J(N_p)$ scale proportionally to  $ N_p^{-1/2}$ for integrations comprising $1$ to $10^5$ pulses. 
The levels of jitter are consistent in observations at different epochs.
At $1400$~MHz, we find that jitter contributes $\approx 40$~ns (rms)  to the residuals for an  hour-duration observation, which is consistent with previous estimates for the level of jitter noise for this pulsar \cite[][]{2012MNRAS.420..361L}.

The level of jitter noise  is modestly smaller at higher frequency and modestly greater at lower frequencies.
 This is likely related to  the narrowing of the pulse width at higher frequencies.  

Because of the high flux density of the pulsar, all observations in all three bands are jitter-dominated.  It is possible to measure the effects of jitter simultaneously with the 10cm/50cm system and assess the degree of correlation between the bands.
In the upper and middle panels  of figure \ref{fig:resid_corr_J0437}, we show the correlation between residual TOAs formed  from the upper and lower halves of the 10cm and 50cm observations respectively. 
We find that there is a high level of correlation between residuals.   
Within the 50cm band we find that the correlation coefficient is $0.7$.  The probability of the null hypothesis (that there is no correlation)  is $3\times10^{-17}$, indicating that the correlation is highly significant.
Within the 10cm band we find that the  correlation coefficient is $0.8$. The probability of the null hypothesis is $1\times 10^{-15}$, again indicating that the correlation is highly significant. 
In the bottom  panel of Figure \ref{fig:resid_corr_J0437}, we plot residuals formed from the nearly simulatneous observations with the 10cm and 50cm system.
The start times for the observations were different in the bands by $1$~s, which should decorrelate the residuals TOAs by only a small amount because profiles are formed from $60$~s sub-integrations.
 The correlation coefficient for the residuals TOAs in both bands is found to be $0.2$,.  The probability of no correlation is also $0.2$, indicating that there is no evidence for correlated TOAs. We therefore  place a limit on the jitter bandwidth of $\lesssim 2$~GHz. 
We attempted to form TOAs in finer sub-bands of the $3100$~MHz system, in order to assess if there is a loss of correlation over the $1024$~MHz band of the system.
We find no evidence for a decorrelation.


\subsection{PSR~J1022$+$1001}

\begin{figure}
\begin{center}
\includegraphics[scale=0.5]{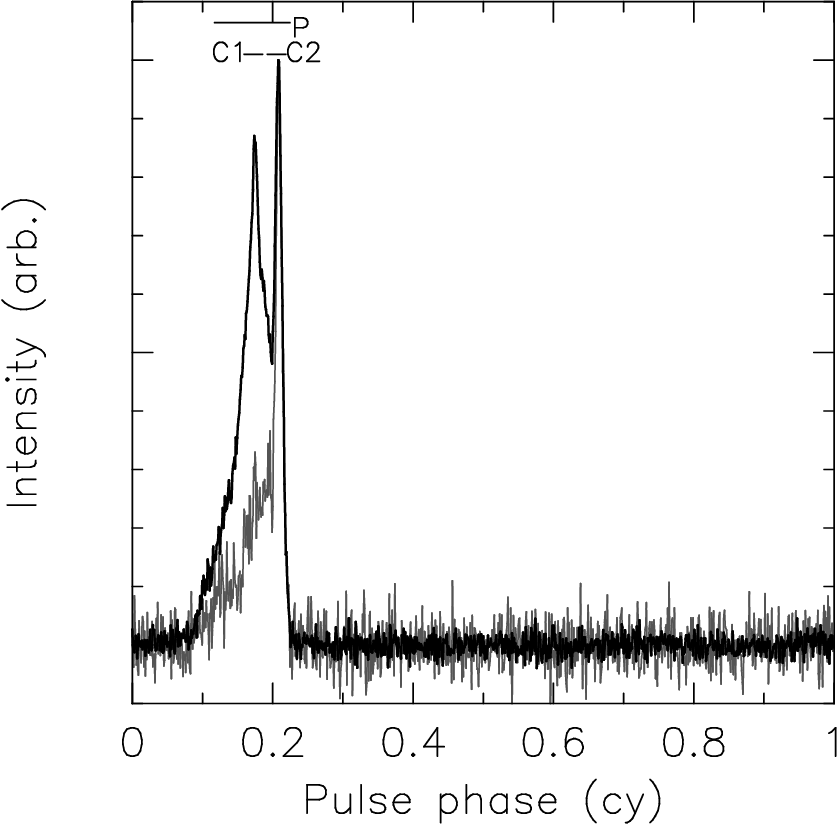}  \\
\caption{\label{fig:profiles_1022} Pulse profiles for PSR~J1022$+$1001.  The thick line shows the  average pulse profile formed from all of our single-pulse observations.   The thinner grey line shows the average of the $100$ brightest pulses.    The profiles have been normalised to have the same peak flux density.}
\end{center}
\end{figure}

\begin{figure}
\begin{center} 
\includegraphics[scale=0.4]{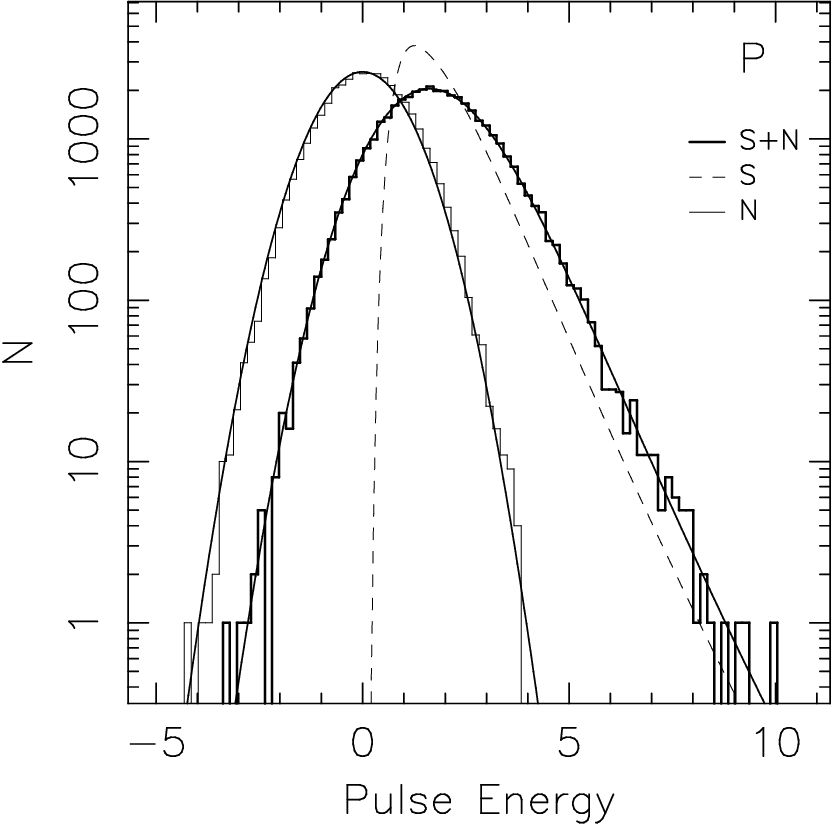} \\
\vspace{2mm}
\includegraphics[scale=0.4]{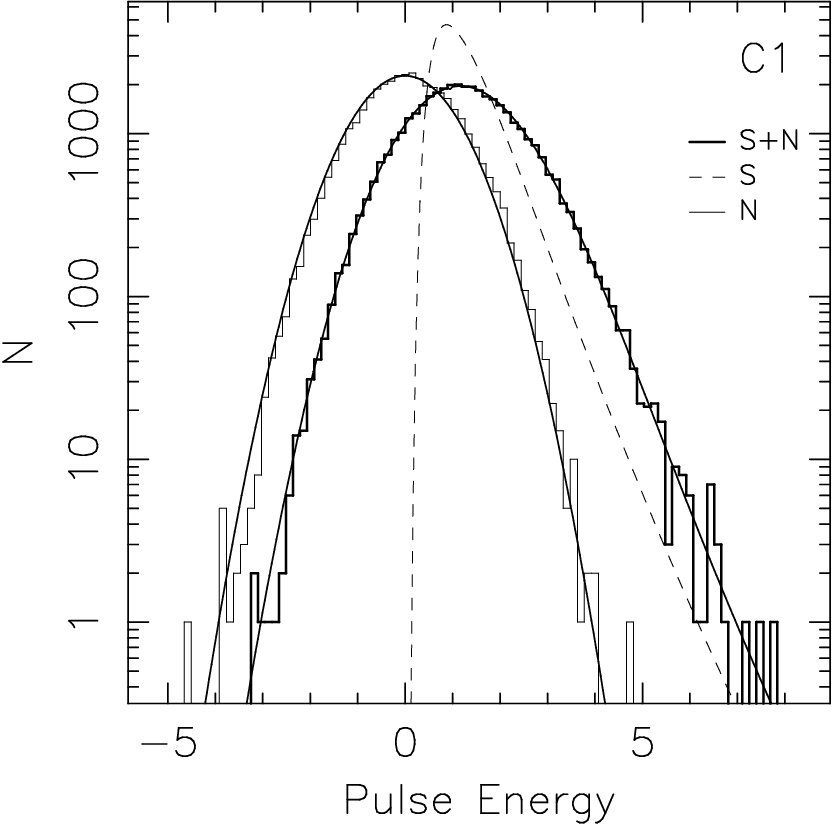}  \\
\vspace{2mm}
\includegraphics[scale=0.4]{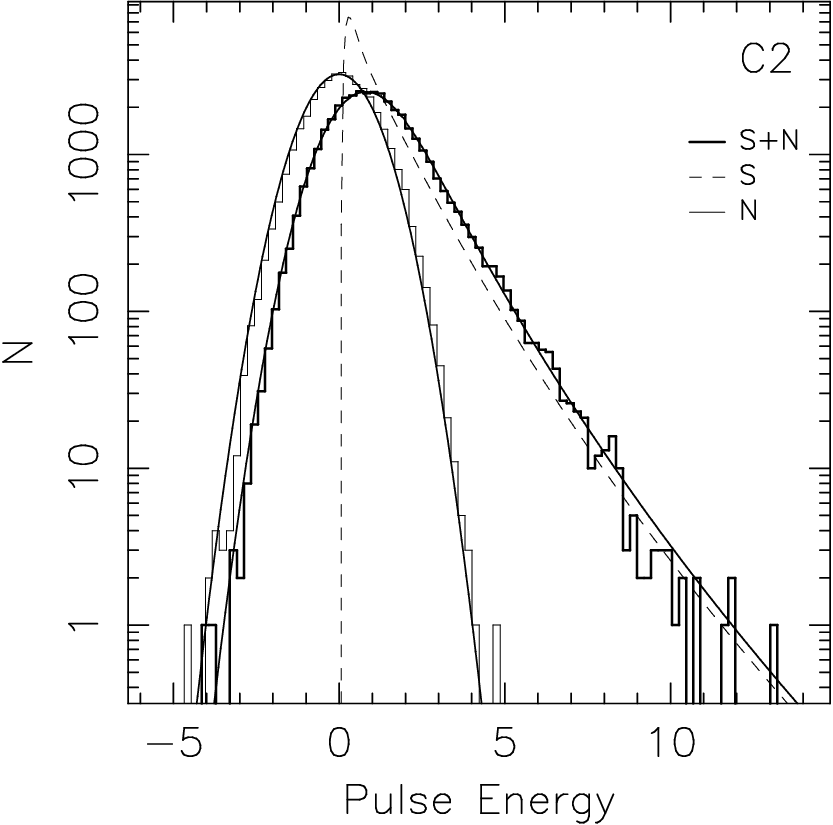} \\
\caption{  \label{fig:energy_hist_J1022}  \footnotesize    Pulse-energy distributions for PSR~J1022$+$1001.  In the uppermost panel, we show the energy distribution  over the majority of the  pulse (window P in Figure \ref{fig:profiles_1022}), in the middle panel we show the energy distribution for a window C1 centred on the leading component,  and in lowermost panel we show the energy distribution in a window C2 centred on the trailing component.     The labels are the same as in Figure \ref{fig:energy_hist_J1713}.    }  
\end{center}
\end{figure}

PSR J1022$+$1002 is a relatively bright ($6$~mJy phase-averaged flux density)   pulsar with a $16$~ms spin period.  It scintillates strongly at $1400$~MHz, enabling studies of pulse-shape changes and short-term timing variations with the Parkes telescope.
Previous  timing analysis at both   Parkes    and other observatories  show that  PSR J1022$+$1002 has  timing variations well in excess of those expected from radiometer noise.  This excess has been attributed to long-term pulse profile instabilities \cite[][]{1999ApJ...520..324K} and imperfect polarisation calibration \cite[][]{2004MNRAS.355..941H,2013ApJS..204...13V}.  Here we identify a component of this excess associated with single-pulse variability.


At $1400$~MHz, the profile of PSR~J1022$+$1001, displayed in  Figure \ref{fig:profiles_1022}, is dominated by two components.
In our analysis of pulse energy we measured the pulse energy from a window encompassing most of the main pulse (labelled P in Figure \ref{fig:profiles_1022}) and windows centred on the dominant leading (labelled C1) and trailing  (labelled C2) components.  
 The dominant components have approximately the same total intensity.   The trailing component is nearly $100 \%$ linearly polarised, whereas the leading component shows relatively low levels of polarisation.
The pulse profile formed from the  brightest $100$  pulses is also displayed in Figure \ref{fig:profiles_1022}. 
While bright individual pulses are found to be centred on both components C1 and C2, the majority of the brightest pulses originate from C2.   In the profile formed from the brightest pulses,  component C1 is  approximately five times weaker than  component  C2.

The pulse-energy distribution for PSR~J1022$+$1001, and its subcomponents,  displayed in   Figure \ref{fig:energy_hist_J1022}, show log-normal distributions.
 The best-fitting models for the pulse-energy distributions are listed in Table \ref{tab:energy_models}.    
The window around the leading component C1 has a larger mean energy, but a lower intensity modulation than the window around component C2, consistent with observation that the bright pulses are dominated by the second component. 
We find no evidence for correlations in the energies of components C1 and C2. 


When calculating the levels of jitter noise, we find that $\sigma_J(N_p) \propto N_p^{-1/2}$  with consistent levels of jitter noise inferred from different backends.
  In $1$~hr of observations we estimate that jitter noise contributes $\approx 280$~ns rms noise to the observations at $1400$~MHz.   At $3100$~MHz we find comparable levels of pulse jitter noise to that measured at $1400$~MHz.  At $730$~MHz, we find that the level of pulse jitter was less than that measured at higher frequency.

Polarisation calibration  and pulse profile evolution also cause measurable pulse-shape changes for this pulsar, and have previously  limited its timing precision.
Because of the high level of linear polarisation of narrow component C2, PSR~J1022$+$1001 is especially susceptible to polarisation calibration errors.
In an analysis of $5$~yr of observations of this pulsar \cite{2013ApJS..204...13V}  showed that improper polarisation could contribute $\sim 800$~ns of excess (rms) scatter to the residual arrival times. 
The pulsar shows significant pulse profile evolution, with the leading component C1  having a large  spectral index.
At $1400$~MHz, the pulsar scintillates strongly, and the combination of pulse profile evolution and scintillation can cause  significant variations in the frequency-averaged pulse profile. 
Both of these effects are correctable; if these effects are corrected jitter noise will limit the achievable timing precision for this pulsar.

\subsection{PSR~J1603$-$7202}

\begin{figure}
\begin{center}
\includegraphics[scale=0.5]{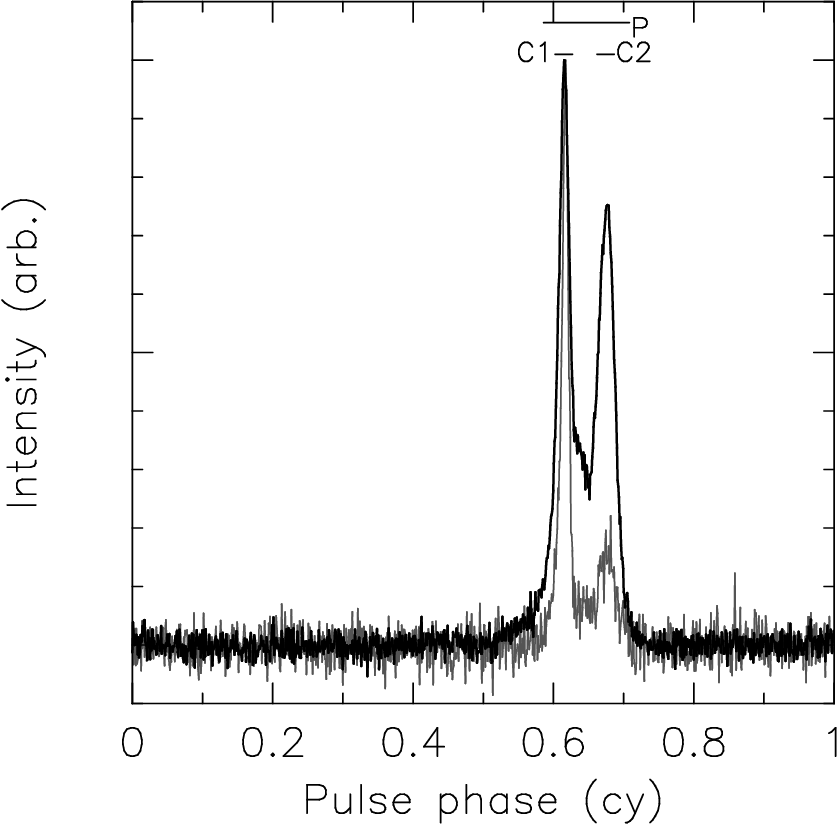}  \\
\caption{\label{fig:profiles_1603} Pulse profiles for PSR~J1603$-$7202.  The thick line shows the  average pulse profile formed from all of our single-pulse observations.   The thinner grey line shows the average of the $100$ brightest pulses.   The profiles have been normalised to have the same peak flux density.}
\end{center}
\end{figure}

PSR~J1603$-$7202 has a spin period of $\sim 15$~ms.   
At $1400$~MHz, the profile of PSR~J1603$-$7202, displayed in  Figure \ref{fig:profiles_1603}, is dominated by two components  connected by a bridge of emission. We measured pulse energies in a window containing most of the main pulse (labelled P in Figure \ref{fig:profiles_1603}), and smaller windows centred on two dominant sub-components (labelled C1 and C2).   The pulse profile formed from the $100$~brightest pulses is also displayed in  Figure \ref{fig:profiles_1603}. 
While the brightest individual pulse, integrated over the full window,  was dominated by the trailing component C2, the vast majority of the bright  pulses, and the brightest pulses in narrower windows tended to originate from component C1.  


\begin{figure}
\begin{center} 
\includegraphics[scale=0.4]{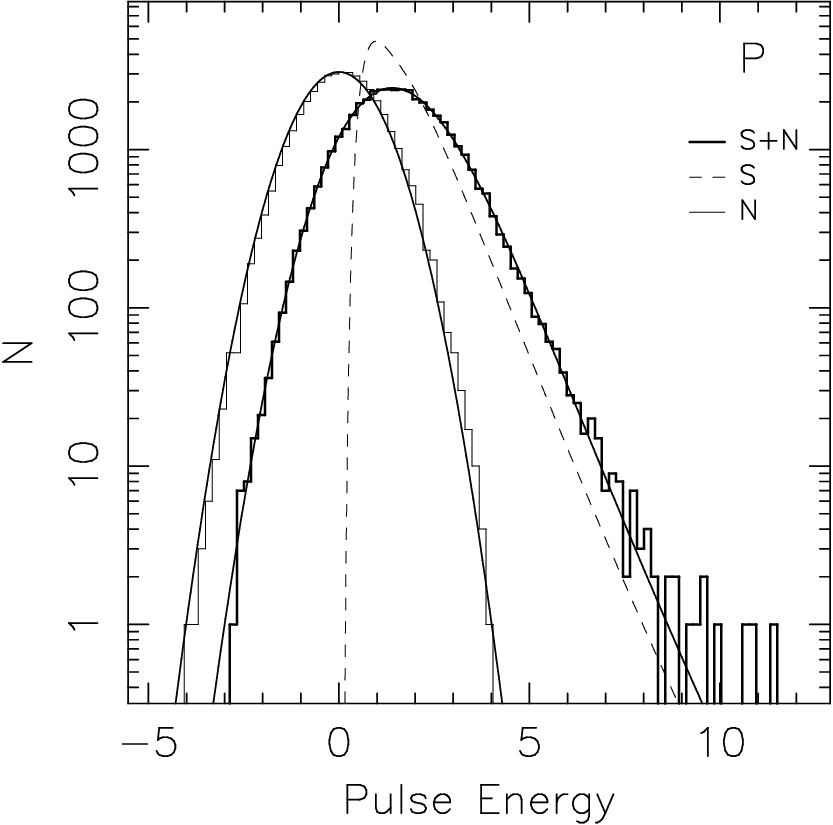} \\
\vspace{2mm}
\includegraphics[scale=0.4]{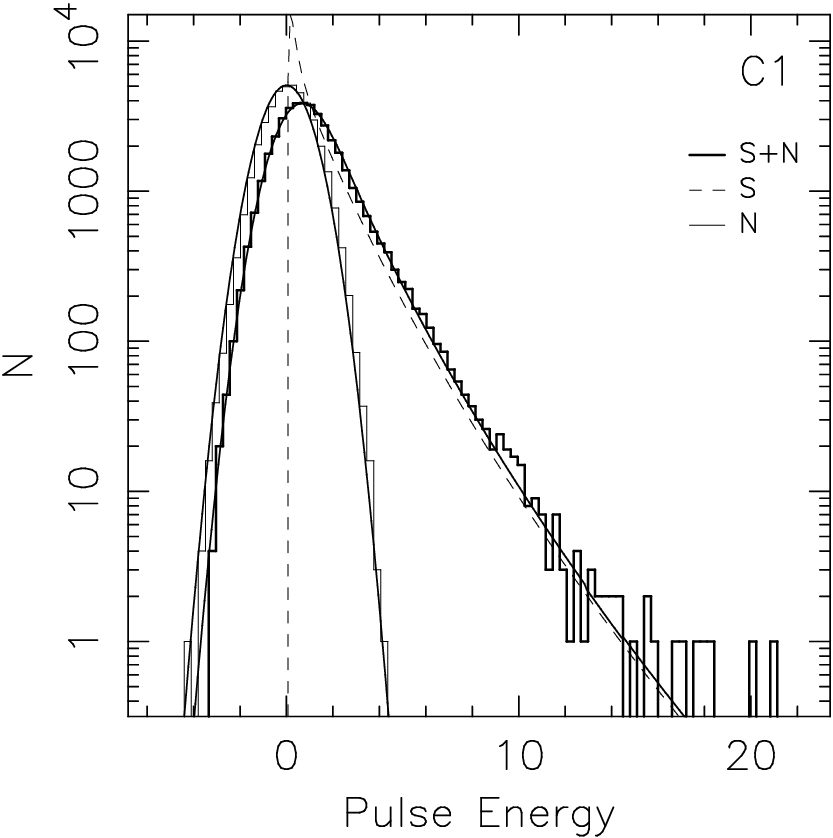} \\
\vspace{2mm}
\includegraphics[scale=0.4]{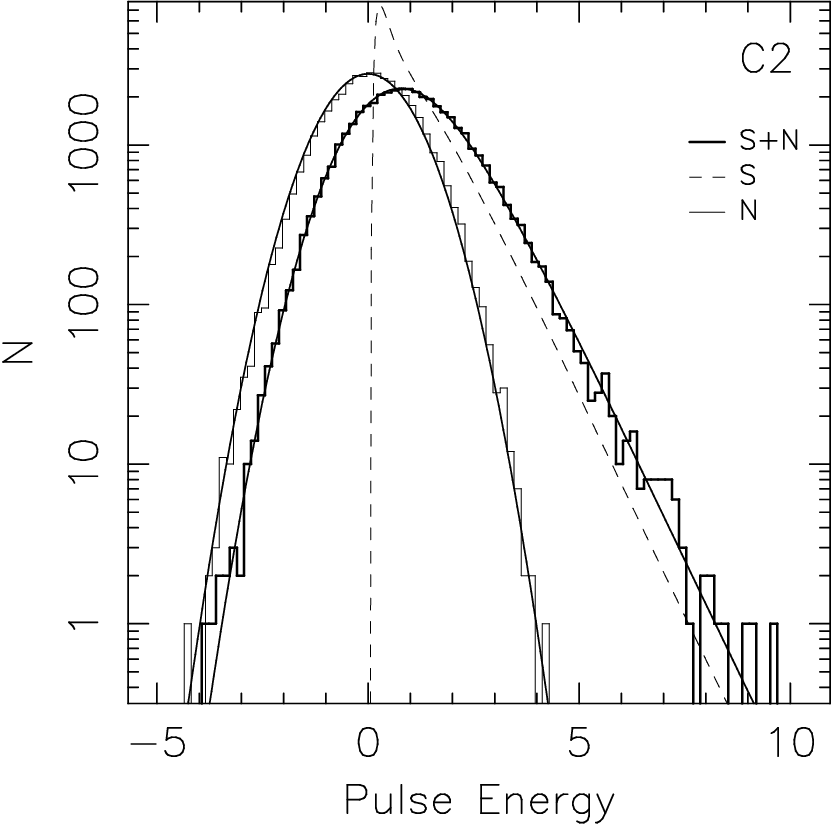} \\
\caption{  \label{fig:energy_hist_J1603}  \footnotesize  Pulse-energy histograms for PSR~J1603$-$7202.  In the uppermost panel, we show the pulse energy measured  in window P containing  most of the pulse profile, as identified in the uppermost panel of Figure \ref{fig:profiles_1603}.  In the middle and lowermost panels, we show the pulse energy measured in windows C1 and C2, also identified in Figure \ref{fig:profiles_1603}, which are centred, respectively, on the two leading and trailing components of the pulse profile.     The labels are the same as in Figure \ref{fig:energy_hist_J1713}.   }  
\end{center}
\end{figure}

The pulse-energy distribution, displayed in Figure \ref{fig:energy_hist_J1603}, shows evidence for approximately log-normal statistics over in the  windows containing the  main pulse and components C1 and C2. 
 Window C1, containing the leading component,  has a lower mean energy, but a higher variance (and hence higher modulation index) than window  C2, which contains the  trailing component.  
  The best-fitting models for the pulse-energy distributions are presented in Table \ref{tab:energy_models}. 
We find no evidence for correlation between the components C1 and C2.

When calculating the level of jitter noise, we find $ \sigma_J(N_p) \propto N_p^{-1/2}$ and we estimate that in a $1$~hr observation at $1400$~MHz that jitter induces an rms error of $\approx 200$~ns.  We were unable to detect the presence of jitter noise at other frequencies, but the upper limits were consistent with the $1400$~MHz observations.


\subsection{PSR~J1744$-$1134}

\begin{figure}
\begin{center} 
\includegraphics[scale=0.5]{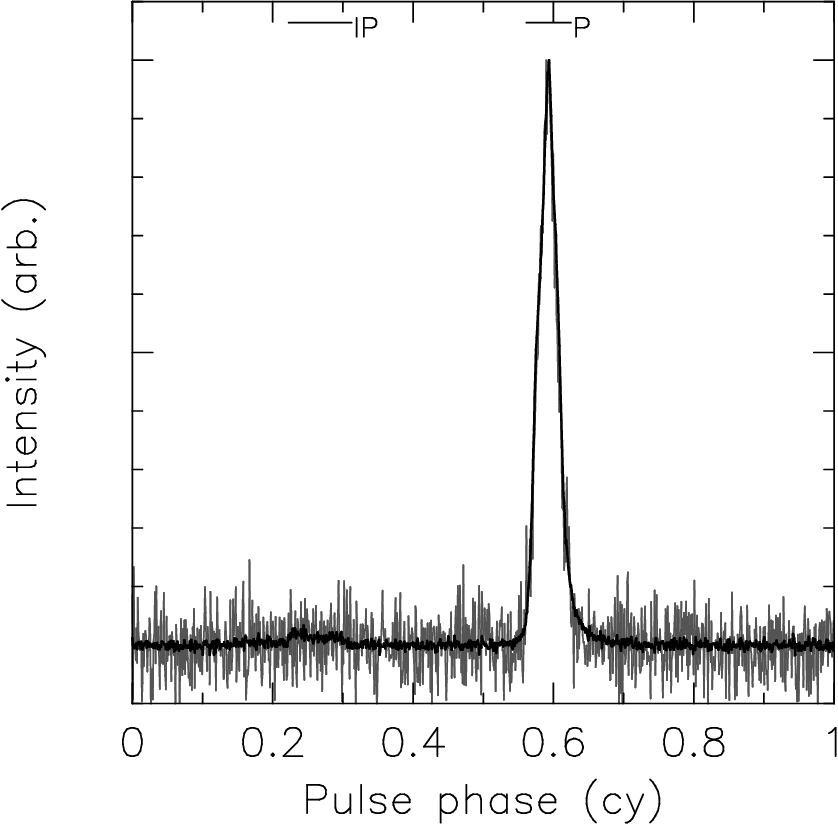} \\
\vspace{2mm}
\includegraphics[scale=0.5]{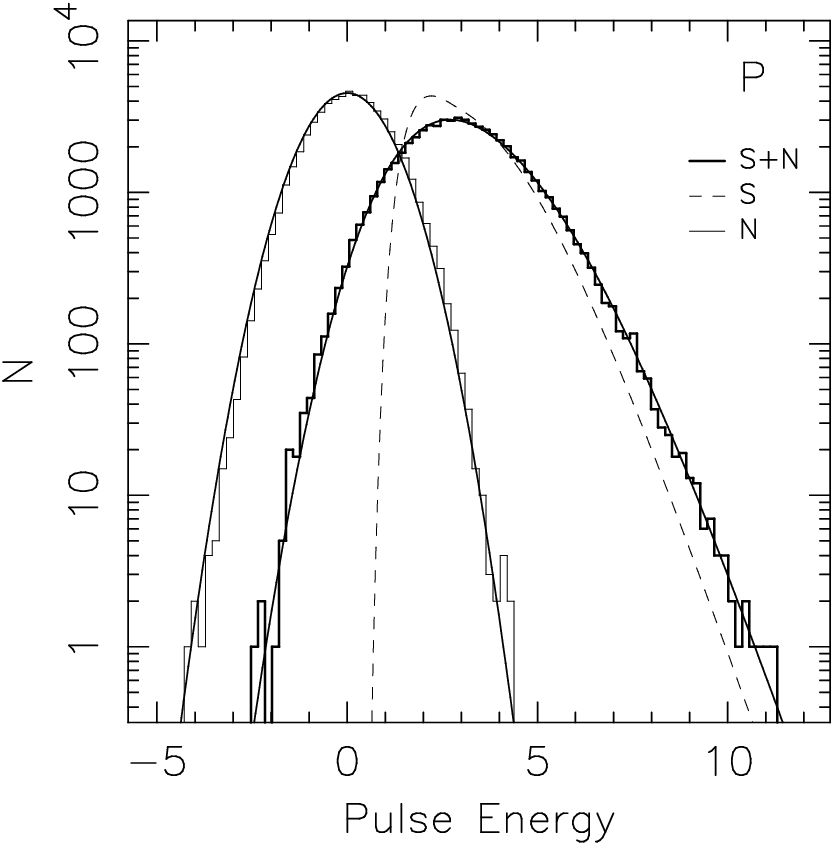} 
\caption{  \label{fig:energy_hist_J1744}  \footnotesize  Top panel:  Pulse profiles for PSR~J1744$-$1134.  The thick line shows the  average pulse profile formed from all of our single-pulse observations.   The thinner grey line shows the average of the $100$ brightest pulses.   The profiles have been normalised to have the same peak flux density.  Lower Panel: Pulse-energy histogram for window P, as displayed in uppermost panel.   The labels for the panel are the same as in Figure \ref{fig:energy_hist_J1713}.    }  
\end{center}
\end{figure}

PSR J1744$-$1134 has a relatively narrow main pulse and a faint interpulse, as displayed in the upper panel of  Figure \ref{fig:energy_hist_J1744}.
The pulsar has a spin period of $\sim 4.1$~ms. While it has a flux density of only $3.1$~mJy, it scintillates strongly  at $1400$~MHz.  

We have identified single-pulse emission from the main pulse  but do not detect  strong pulses from the interpulse. 
The average profile of the $100$ brightest pulses, also displayed in the upper panel of Figure \ref{fig:energy_hist_J1744} is consistent in width with the average of all the pulses, suggesting that bright pulses are emitted over a wide range of pulse phase.
We calculated the pulse-energy distribution in  window P (see Figure \ref{fig:energy_hist_J1744}) containing the majority of the energy of the main pulse.  
We find that  the pulse-energy distribution in this window, displayed in the bottom panel of Figure \ref{fig:energy_hist_J1744},  shows an approximately log-normal energy distribution.

Like the majority of  pulsars in our sample we find that  $\sigma_J(N_p) \propto N_p^{-1/2}$  up to $N_p=6\times10^4$, the largest value we searched.
 Based on these results  we estimate that in $1$~hr of observation jitter noise contributes $\approx 40$~ns rms error to the arrival times for observations close to  $1400$~MHz.
The low levels of jitter noise are attributed to the relatively narrow pulse profile.
For this pulsar we were unable to detect jitter noise in observations at  $3100$~MHz or $730$~MHz due to sensitivity limitations of our observations.  However our upper limits were consistent with the analysis at $1400$~MHz.


\subsection{PSR~J1909$-$3744}

\begin{figure}
\begin{center} 
\includegraphics[scale=0.5]{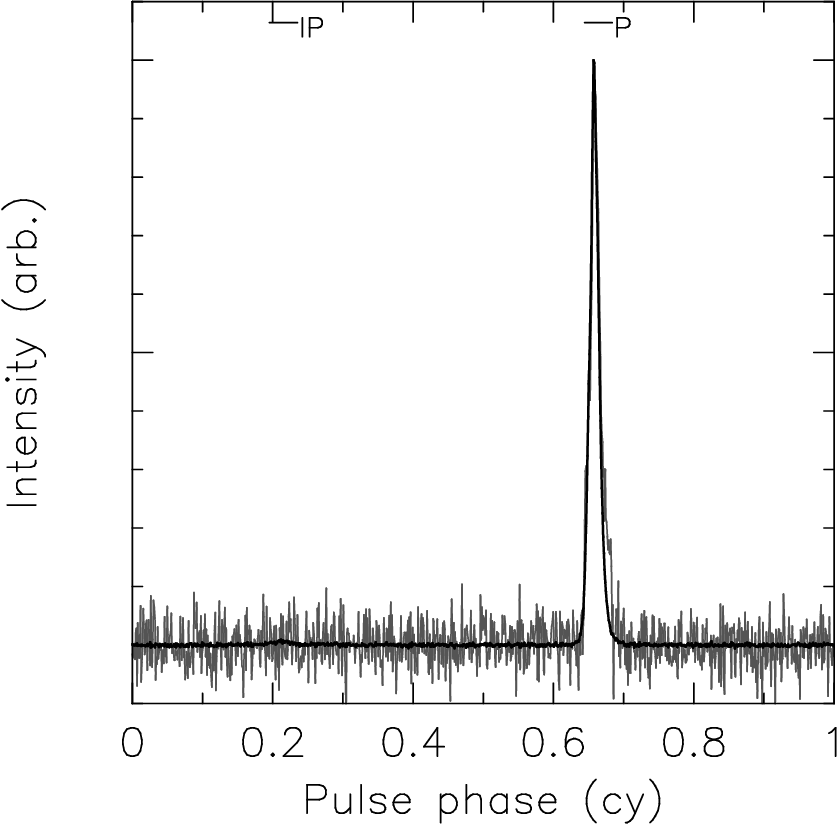} \\
\vspace{0.2in}
\includegraphics[scale=0.5]{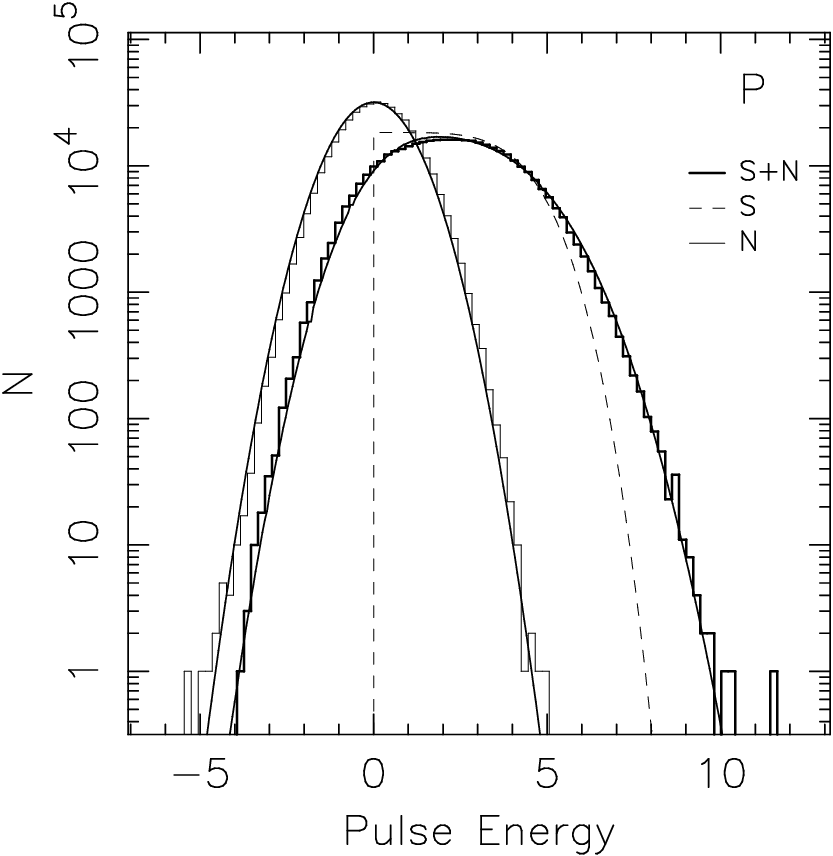} 
%
\caption{  \label{fig:energy_hist_J1909}  \footnotesize   Upper panel: Pulse profiles for PSR~J1909$-$3744.  The thick line shows the  average pulse profile formed from all of our single-pulse observations.   The thinner grey line shows the average of the $100$ brightest pulses.   The profiles have been normalised to have the same peak flux density. Lower panel:   Pulse-energy distribution in window P for  PSR~J1909$-$3744. The labels are the same as in Figure \ref{fig:energy_hist_J1713}.      }  
\end{center}
\end{figure}

\begin{figure}
\begin{center} 
\includegraphics[scale=0.5]{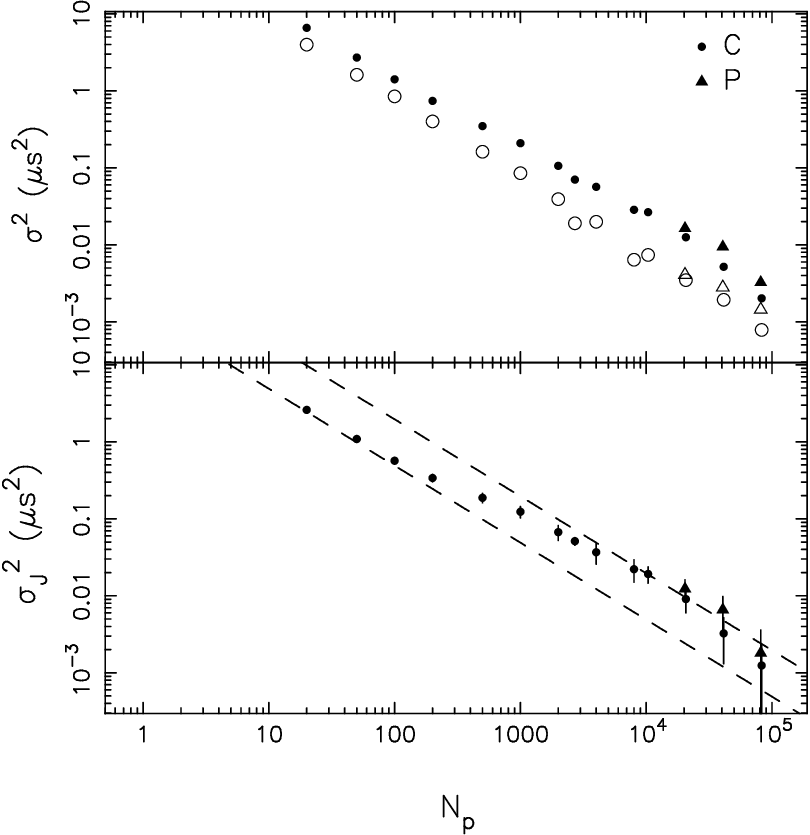}
\caption{  \label{fig:emission_noise_1909}   Estimates of levels of jitter noise for PSR~J1909$-$3744.  \footnotesize    Upper panel: Variance of residual time series for observed datasets (filled symbols) and simulated datasets (open symbols).  The symbols shapes represent different backends and are listed below.  Lower panel:  the difference between the observed and simulated variance.  We attribute this difference to pulse-shape variations.   The dashed lines show the levels of jitter noise predicted from observations of $<10^3$ pulses (lower line) and $> 10^4$ pulses (upper line), both scaling $\propto N_p^{-1/2}$.  Symbols:  circles - CASPSR at $1400$~MHz (labelled C); triangles -  PDFB3 at $1400$~MHz (labelled P).}  
\end{center}
\end{figure}

\begin{figure}
\begin{center} 
\includegraphics[scale=0.6]{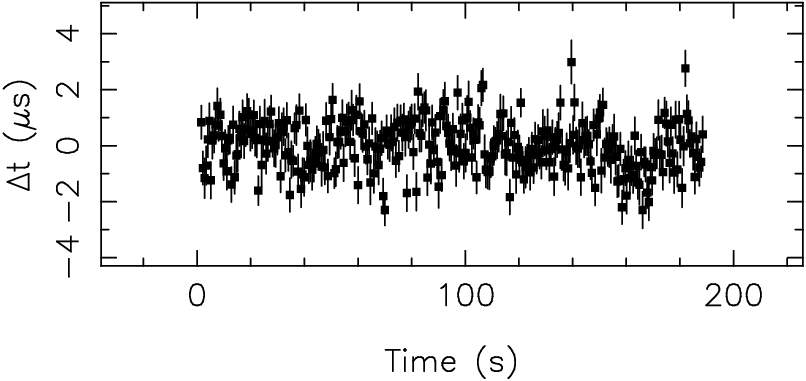} \\
\includegraphics[scale=0.6]{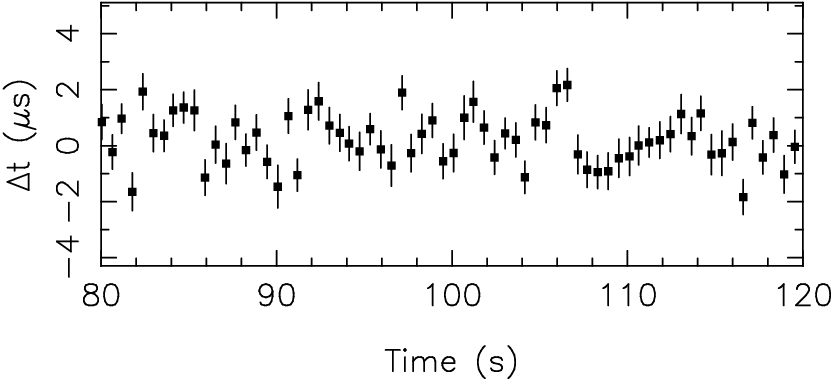}
\caption{  \label{fig:J1909_residuals_short}   Top panel: Residual TOAs for PSR J1909-3744, derived from pulse profiles formed from $200$ consecutive pulses.  Bottom panel:  Sub-interval that highlights correlated residuals.   }  
\end{center}
\end{figure}

At decimetre wavelengths, PSR J1909$-$3744 shows a narrow $42~\mu$s~wide  main component and a faint interpulse, with both  identified in the upper panel of Figure \ref{fig:energy_hist_J1909}.  
Single pulses were detected from the main pulse but not from the interpulse. 
The average pulse profile formed from the $100$ brightest pulses is also displayed in the upper panel of Figure \ref{fig:energy_hist_J1909}. 
The profile width is $\approx 80\%$ of the width of the profile of all the pulses, indicating  that, like PSR~J1744$-$1134, and unlike PSR~J0437$-$4715,  bright single pulses are emitted across nearly the entire width of the main pulse.

Of all the pulsars in the sample, PSR~J1909$-$3744 deviates the most from a log-normal distribution. The pulse-energy distribution more closely resembles a Gaussian distribution.    The energy distribution for a window centred on the main pulse (labelled P in Figure \ref{fig:energy_hist_J1909}  and its  best-fitting generalised Gaussian model are displayed in the bottom panel of Figure \ref{fig:energy_hist_J1909}.  Relative to a Gaussian distribution ($\alpha = 2$ in Equation \ref{eqn:gaussian}), the pulse-energy distribution  shows a broader distribution about the mean value (i.e., is platykurtic).


 This lack of bright pulses contributes to the low levels of jitter noise for the pulsar.  The paucity of  bright $>5\sigma$ pulses results in small (but measurable) levels of pulse distortion.

Unlike other pulsars in the sample, we find evidence that $\sigma_J(N_p)$ does not scale with a single power law  $\propto N_p^{-1/2}$ as would be expected if no temporal correlations between pulses exist. 
In Figure \ref{fig:emission_noise_1909}, we show how $\sigma_{\rm obs}$ and $\sigma_J$ scale with $N_p$. 
For profiles averaged  from much less than $10^3$ pulses, and profiles averaged from much longer than $10^3$ pulses show $\sigma_J \propto N_p^{-1/2}$, but offset from each other.
We searched for periodicities in the pulse energy using two-dimensional fluctuation spectra.  We observed excess power at low (but nonzero) fluctuation frequency,  but did not find any evidence for periodic features. 
In Figure \ref{fig:J1909_residuals_short}, we show residual TOAs formed from averages of $200$ pulses.  The TOAs show variations that are correlated over $2$~s ($\approx 2000$ pulses), much shorter than the time scale of the structure observed in PSR~J1713$+$0747.  Power spectral analysis of the residuals shows the presence of power at low fluctuation frequency;  however there is no evidence for significant periodicities.

We estimate that at $1400$~MHz, jitter noise contributes approximately $10$~ns rms timing error for an hour-long  observation, the lowest level of any pulsar in our sample.
We were unable to detect the presence of jitter at $3100$~MHz or $730$~MHz, but the limits on the level of jitter were consistent with observations at other frequencies.

\subsection{PSR~J2145$-$0750}

\begin{figure}
\begin{center}
\includegraphics[scale=0.5]{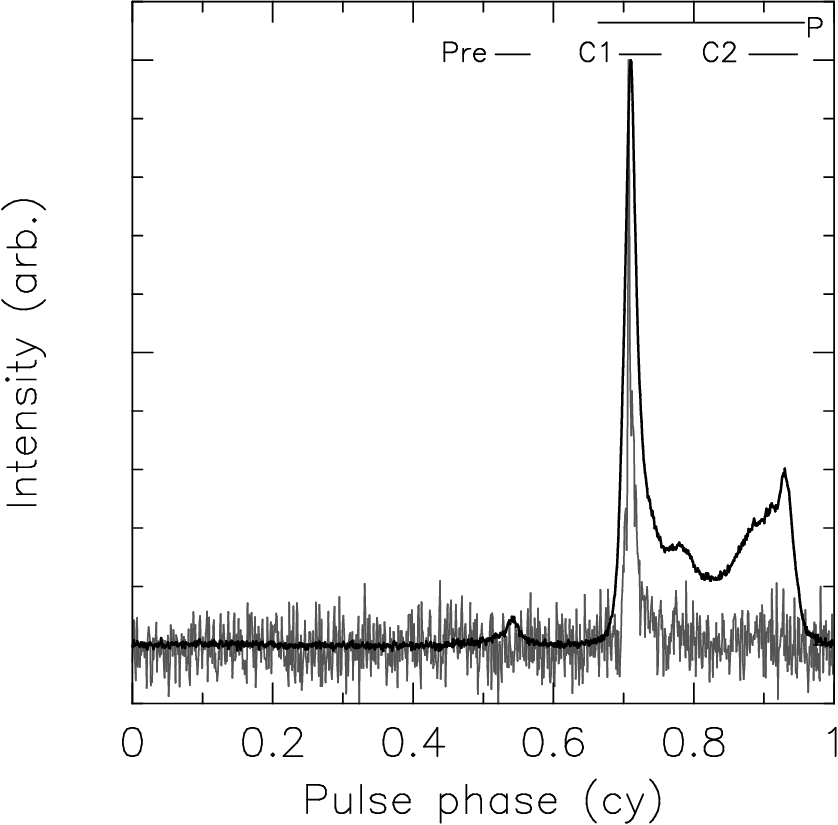}  \\
\caption{\label{fig:profile_J2145} Pulse profiles for PSR~J2145$-$0750.  The thick line shows the  average pulse profile formed from all of our single-pulse observations.   The thinner grey line shows the average of the $100$ brightest pulses.  The profiles have been normalised to have the same peak flux density. }
\end{center}
\end{figure}

\begin{figure}
\begin{center} 
\includegraphics[scale=0.4]{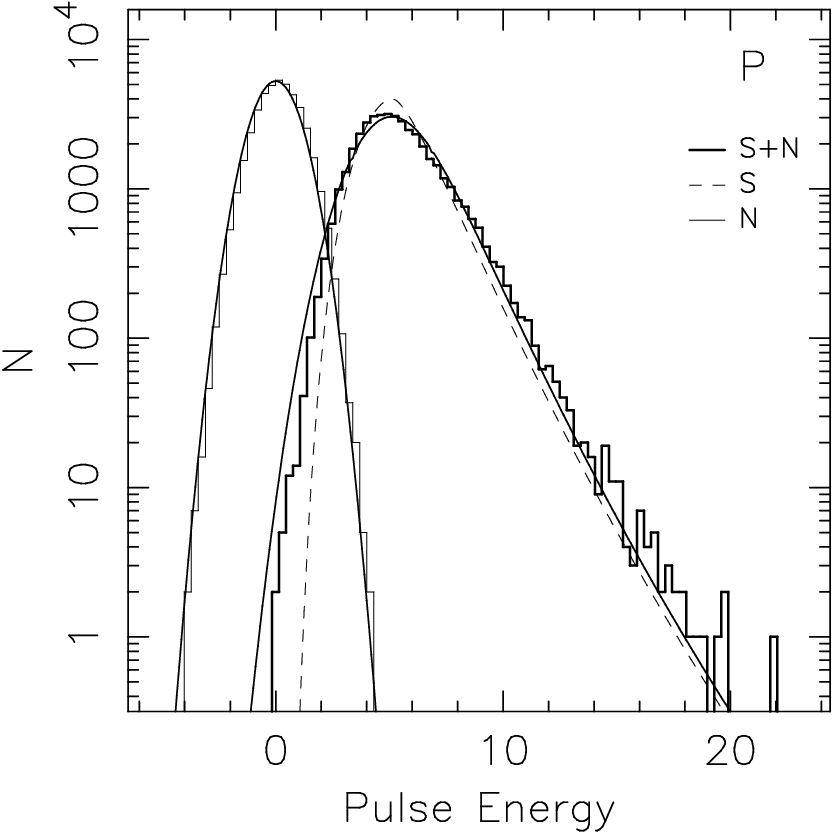} \\
\vspace{2mm}
\includegraphics[scale=0.4]{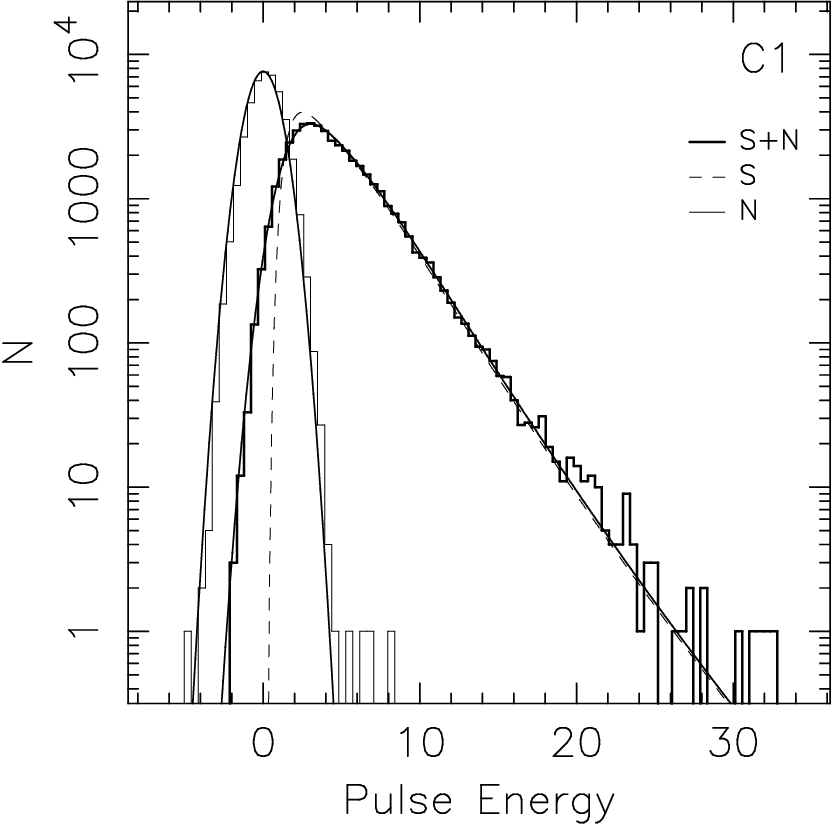}  \\
\vspace{2mm}
\includegraphics[scale=0.4]{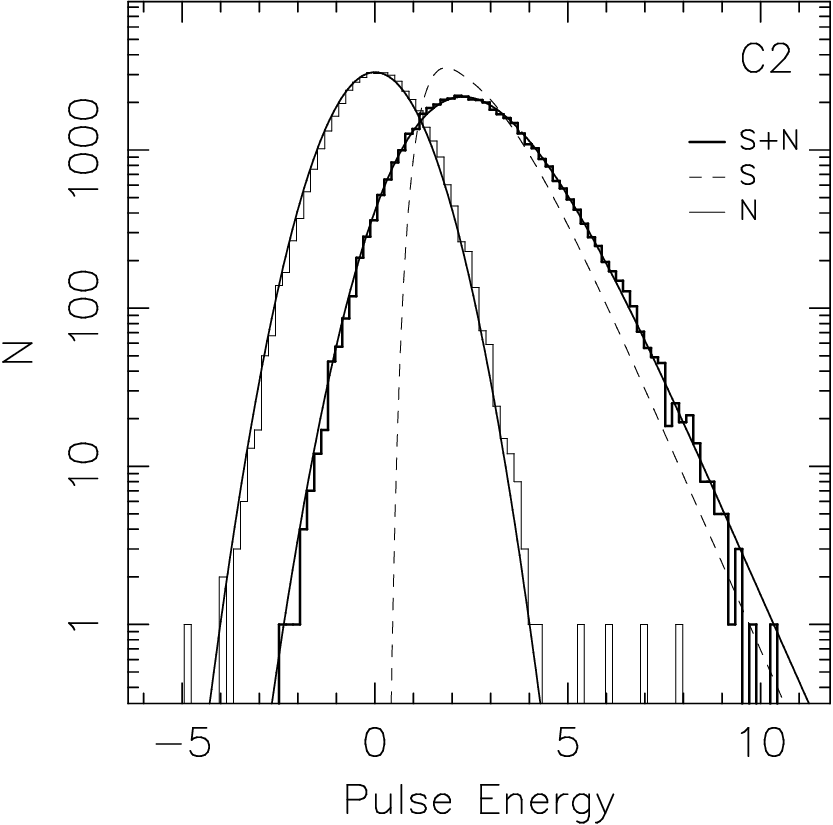} \\
\caption{  \label{fig:energy_hist_J2145}  \footnotesize  Pulse-energy histograms for PSR~J2145$-$0750.  In the uppermost panel, we show the pulse-energy distribution for the main pulse, identified as region P in the upper panel of Figure \ref{fig:profile_J2145}.  In the centre and lowermost panels we show respectively the pulse-energy distribution in windows C1 and C2,  also labelled in Figure \ref{fig:profile_J2145}, centred on components C1 and C2.     See Figure \ref{fig:energy_hist_J1713} for a description of the plot. } 
\end{center}
\end{figure}

\begin{figure}
\begin{center} 
\includegraphics[scale=0.4]{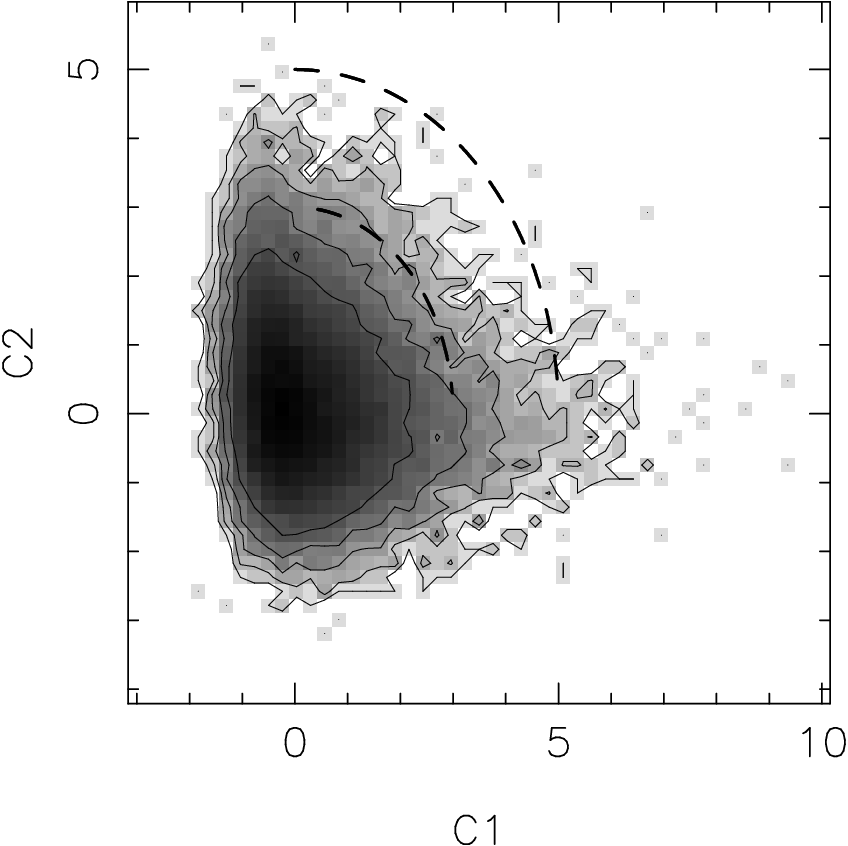}
\caption{  \label{fig:energy_corr_J2145}  \footnotesize Correlations in  energy of components C1 and C2 for PSR~J2145$-$0750.  The components have been normalised to have unit mean and variance.  The dashed lines shows the ($1-\sigma$) and ($2-\sigma$) contour of equal probability if component energies are independent.    }  
\end{center}
\end{figure}

 PSR~J2145$-$0750 has a spin period of $\sim 15$~ms and, at  $1400$~MHz,  the pulsar shows a complex profile morphology,  with two strong components  (C1 and C2) and a precursor (Pre), as displayed   in Figure   \ref{fig:profile_J2145}.  
We  have identified strong pulses centred on both main features.
We did not detect any  bright pulses from the precursor. 
A pulse profile formed by adding the $100$ brightest pulses, displayed in Figure \ref{fig:profile_J2145}, shows that the brightest pulses generally originate from the leading component C1.

PSR J2145$-$0740 shows similar pulse-energy characteristics to other multi-component pulsars in our sample.
The pulse-energy distributions, plotted in Figure \ref{fig:energy_hist_J2145}, show log-normal statistics.
The best-fitting model distributions are displayed in Table \ref{tab:energy_models}. 
We find that the  energy distribution in a window C1 around the  leading component has  larger mean and modulation than the window C2 around the trailing component.  
in Figure \ref{fig:energy_corr_J2145}, we show   the  joint distribution of pulse energies for  in windows C1 and C2. 
Because the distributions of the energies are non-Gaussian we use the non-parametric Spearman rank-order correlation coefficient to test the level of correlation.
The Spearman correlation coefficient was found to be $-0.2$,  The probability of the null hypothesis  (that there is no correlation) was calculated to be $8\times 10^{-11}$, indicating that negative correlation is highly significant.


The variable pulse morphology introduces large levels of jitter noise. 
We find that $\sigma_J \propto N_p^{-1/2}$ for $N_p < 1.5\times10^4$, the largest value we searched.     
Because of the large pulse width and slow pulse period, the inferred levels of jitter noise are large, contributing $\approx 190$~ns rms timing error for  hour-long  observations at $1400$~MHz.  We were able to detect the presence of jitter noise at $3100$~MHz and $730$~MHz as well.  Unlike the other pulsars in the sample the level of jitter noise increases at higher frequency, with the estimated jitter noise largest at $3100$~MHz, as displayed in Table \ref{tab:emission_noise}.  At $1400$~MHz and $730$~MHz, the jitter levels for the pulsar are comparable with each other. 


\subsection{Other pulsars}

We searched for evidence of jitter in other pulsars in the PPTA sample using only fold-mode observations. 
The analysis was identical to that of the pulsars discussed above.
However, we had limited range in $N_p$ over which to search for the $\sigma_J \propto N_p^{-1/2}$ scaling expected of jitter.  

We find evidence of S/N-independent noise in the observations of PSR~J1939$+$2134, in observations in both bands at $\sim 1400$~MHz and $730$~MHz.   
In the $1400$~MHz band, we associate this with jitter noise.  At $730$~MHz, we associate this noise with variable (and stochastic) broadening of the pulse profile, referred to as the finite-scintle effect by \cite{cs2010}.    Indeed, the effects of stochastic pulse broadening have been measured for this pulsar  at $430$~MHz \cite[][]{1990ApJ...349..245C,2011MNRAS.416.2821D}.

At low frequencies, the pulse profile is broadened by multi-path propagation through the interstellar medium.
The broadening is stochastic, resulting in pulse-shape distortions that affect timing precision.
The rms variations of arrival times of \cite[][]{cs2010} induced by stochastic broadening is
 \begin{equation}
\sigma_{\rm DISS} = C_1 \frac{\tau_d}{N_s},
\end{equation}
where $\tau_d$ is the pulse broadening time $N_s$ is the number of scintles in the observation
\be
\label{eqn:ns}
N_s = \left (1 + \eta \frac{\Delta  T}{t_d} \right) \left(   1  + \eta \frac{\Delta \nu}{\nu _d }\right),
\ee 
and $C_1$ is a constant of order unity.
In Equation (\ref{eqn:ns}), $\Delta T$ is the observing time, $\Delta \nu$ is the observing bandwidth, $\nu_d$ is the diffractive interstellar scintillation (DISS) scale, $\Delta t_d$ is the diffractive time scale, and $\eta$ is the filling factor of the scintles.  Following convention,  $\Delta  \nu_d$ is the half width at half maximum, and $\Delta t_d$ is the half width at the $1/e$ point.
\cite{1985ApJ...288..221C} found that  $\eta \sim 0.2$ and we will assume that value.

In the $1400$~MHz band, we find that there is an excess noise with rms amplitude of $40$~ns in $T=30$~s sub-integrations, with a bandwidth of $300$~MHz.
The noise was observed to be uncorrelated from sub-integration to sub-integration.
To estimate the effects of scintillation we calculated  the dynamic spectrum of the observation and then formed  its  two-dimensional autocorrelation function (ACF),  measuring  its decorrelation time to be $t_d \approx 380$~s, and decorrelation bandwidth to be $\nu_d \approx 1.2$~MHz.  
Based on these values, we expect that in $30$~s sub-integrations,  stochastic broadening of the pulse profile  induces $\sim 20$~ns of RMS error, which is a factor of $2$ smaller than what is measured.  
The measured noise is consistent with jitter noise observed in other pulsars, in that the inferred jitter parameter (discussed further below)  is in range of values measured for other pulsars in the sample.
Higher time resolution observations could be used to distinguish intrinsic shape variations from DISS effects.

There is stronger evidence that our observations at $730$~MHz are  limited by stochastic broadening.
We find that there is time-correlated structure in the residuals that contributes rms scatter of $140$~ns to the observations that are correlated over $\sim 200$~s, in sub-integrations of $30$~s duration.
In an analysis of the ACF of the dynamic spectrum, we found  $t_d \approx 150$~s and $\nu_d \approx 53$~kHz.   
We therefore expect stochastic pulse broadening to contribute $\sim 180$~ns rms to observations, which is only $30\%$ larger than the measured values.

We do not expect DISS to play a role in any of the pulsars for which we have detected jitter noise.  These pulsars are much more weakly scattered, and even in the 50cm band, the contribution from scattering is expected to be $< 10$~ns \cite[][]{2010ApJ...717.1206C}.

In the remaining pulsars in the sample, we attribute the non-detection of jitter noise to the low flux densities of the pulsars. 
The effects of jitter are expected to be significant when the instantaneous S/N exceeds unity.
We place conservative limits on the level of jitter of the other pulsars by assuming that the jitter contribution to the TOA error is smaller than the total observed rms of the residuals: 
\be
\sigma_J(N_p) < \sigma_{\rm obs} (N_p).
\ee

In Table \ref{tab:emission_noise}, we present measurements of, or limits on, the level of 
jitter noise in PPTA pulsars.  We show both the noise expected at the single-pulse level $\sigma_J(1)$ and in $1$~hr of observation $\sigma_J({\rm hr})$ in addition to the  jitter parameter calculated using $W_{\rm eff}$,  $W_{50}$, and $W_{\rm eff}(1+m_I^2)$ as measurements of pulse widths.

\begin{figure}
\begin{center}
\includegraphics[scale=0.4]{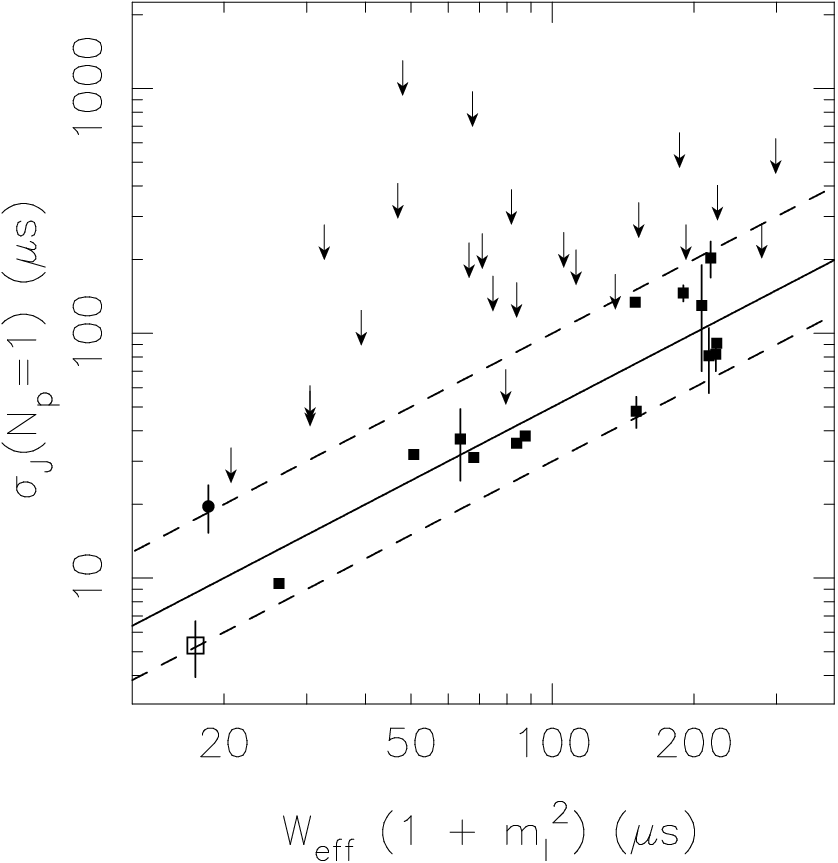}
\caption{  \label{fig:cor_jit}   Correlation of pulse width $W_{\rm eff} (1 + m_I^2)$  and level of jitter noise $\sigma_J(N_p=1)$.   The downward pointing arrows show upper limits on the levels of jitter noise.  The lines represent different models $\sigma_J = \bar{f}_J W_{50}$, where $\bar{f}_J$ is the mean value for the relationship.   The solid line is the best-fitting value $ \bar{f}_{J}= 0.5 \pm 0.1$, the lower and upper dashed lines show the trends $\bar{f_J} = 0.3$ and $\bar{f}_J=1.0$ respectively.   The  open square  show measurements  of jitter noise for PSR J1939$+$2134 at $1400$~MHz, while the  filled circle shows excess noise at  $730$~MHz that we attribute to the effects of stochastic pulse broadening.  }  
\end{center}
\end{figure}

We measured the correlation of $\sigma_J(1)$  with $W_{50}$, $W_{10}$, $W_{\rm eff}$, $W_{50} (1 + m_I^2)$, and $W_{\rm eff} (1+ m_I^2)$ . 
 We find that the strongest correlation is  between $\sigma_J(1)$ and $W_{\rm eff} (1 + m_I)^2$.
   In Figure \ref{fig:cor_jit}, we show the relationship between  between $W_{{\rm eff} } (1+m_I^2)$ and $\sigma_J(1)$.      
We fitted the relationship $\sigma_J(1) = \bar{f}_J  W_{\rm eff} (1+m_{\rm I}^2)$, with $\bar{f}_J$ as a free parameter.
  For pulsars without single-pulse observations, we assume $m_I = 0.3$, though our results were not sensitive to this value.  
We find that the best-fitting jitter parameter is   $\bar{f}_J \approx 0.5$, which is indicated as a solid line Figure \ref{fig:cor_jit}.   
We find that excluding the correction for the modulation index did not change the value $\bar{f}_J$ significantly, and only increased the scatter (and hence decreased the level of correlation). 
A poor correlation is measured between the effective width $W_{{\rm eff}, L}$  (see equation \ref{eqn:effective_width_liu})and $\sigma_J(1)$.  This is not surprising because $W_{{\rm eff}, L}$ is sensitive to broad components that do not contribute greatly to pulse jitter.  


For the other pulsars that we have not detected the presence of jitter noise, we place limits on jitter parameter, and set  upper  limits  $f_J >  0.8$, consistent with the detections of jitter noise and indicating that the non-detections of jitter are associated with insufficient sensitivity. 


\begin{table*}
\caption{ \label{tab:energy_models}  Models for pulse-energy distributions in PPTA pulsars.}
\begin{center}
\begin{tabular}{crrrrlr}
\\ \hline \hline
\multicolumn{7}{c}{Generalised log-normal distribution} \\   \hline
\hline
\multicolumn{1}{c}{PSR}  & \multicolumn{1}{c}{Comp} &  \multicolumn{1}{c}{$  \mu_E $}  &  \multicolumn{1}{c}{$  \ln \sigma_E$} & \multicolumn{1}{c}{ $\alpha$}   & \multicolumn{1}{c}{$m_I$}  & \multicolumn{1}{c}{$\chi^2/N_{\rm DOF}$}  \\ 
\hline
J0437$-$4715 & P & $14.47(1)$ & $0.303(2)$ &   $1.645(7)$ &$0.390(8)$   & $61.6$ \\
                      & C1 & $18.24(2)$ & $0.586(3)$  & $1.884(9)$ & $0.694(2)$  & $51.3$\\
                      & C2 & $3.350(1)$ & $0.197(3)$ & $1.142(8)$ & $0.493(1)$ & $18.3 $ \\
J1022$+$1001 & P & $1.623(1)$ &  $ 0.5023(6)$ & $2.20(9)$ & $0.4859(5)$  & $0.9$ \\ 
                        & C1 & $1.123(3)$ & $0.518(4) $ & $2.19(3)$ & $0.506(3)$  & $0.8$ \\
                       & C2 & $0.794(6)$ & $1.136(7) $ & $2.86(3)$ & $0.958(3)$ & $1.2$ \\
J1603$-$7202 & P & $1.398(2)$ & $0.595(1)$ &  $2.392(5)$ & $0.543(1)$ & $1.5$\\
                       & C1 &   $0.674(1)$ &   $1.420(2)$&$3.161(9)$  & $1.156(1)$ & $1.8$ \\
		     & C2 & $0.728(2)$ & $1.136(2)$ &$3.39(1)$  & $0.852(7)$ & $1.0$ \\
J1713$+$0747 & P & $1.821(1)$ & $0.1715(5)$ & $1.334(2)$  & $0.2870(2)$ &$2.0$  \\ 
J1744$-$1134 & P & $2.836(4)$ & $0.459(6)$ & $2.69(6)$ & $0.388(2)$ & $1.7$\\
J2145$-$0750 & P & $5.276(5)$ & $0.2286(8)$ &  $1.589(6)$ & $0.3003(8)$ & $8.7$ \\ 
 &  C1 & $3.930(5)$ & $0.677(2)$ & $2.44(1)$ & $0.617(1)$  & $1.6$ \\
 &  C2 & $2.376(5)$ &  $0.490(7)$ & $2.45(5)$ & $0.433(2)$  & $0.9$ \\
 \hline
 \multicolumn{7}{c}{Generalised normal distribution } \\   \hline
\multicolumn{1}{c}{PSR}  & \multicolumn{1}{c}{Comp} & \multicolumn{1}{c}{$\mu_E $}&  \multicolumn{1}{c}{$\sigma_E$} & \multicolumn{1}{c}{ $\alpha$} & \multicolumn{1}{c}{$m_I$}   & \multicolumn{1}{c}{$\chi^2/N_{\rm DOF}$} \\ \hline 
J1909$-$3744  & P &  $0.408(3)$ & $4.345(3)$ & $4.275(8)$ & $0.627(2)$  &$33.1$ \\
\hline
\end{tabular}\\
\end{center}
Note:  numbers in parentheses are the uncertainty in the last digit of the parameters, and are derived from Monte Carlo simulation (see text).
\end{table*}

\begin{table*}
\begin{center}
\caption{\label{tab:emission_noise} Jitter noise in PPTA pulsars. }
\begin{tabular}{lrrrrrrcccc}
 \\ \hline  \hline
\multicolumn{1}{c}{PSR}             &  \multicolumn{1}{c}{$\nu$ } &  \multicolumn{1}{c}{$W_{10}$} & \multicolumn{1}{c}{$W_{50}$ }  & \multicolumn{1}{c}{$W_{\rm{eff}}$ } & \multicolumn{1}{c}{$W_{\rm{eff}, M_I}$ }  & $\sigma_{\rm{J}}$(1)     &    $\sigma_{\rm{J}}$(hr)   &   $f_{\rm{J,eff}}$     &     $f_{\rm{J,m_{I}}}$ & $f_{\rm{J,50}}$   \\
             &  \multicolumn{1}{c}{ (MHz)} &  \multicolumn{1}{c}{ ($\mu$s)} & \multicolumn{1}{c}{ ($\mu$s)}  & \multicolumn{1}{c}{ ($\mu$s)} & \multicolumn{1}{c}{ ($\mu$s)}  &  ($\mu$s)     &   (ns)   &       &        \\
\hline
\hline
J0437$-$4715    &   3100      &   267    &   89   &   44 & 51  &   32$\pm$1     &   41$\pm$2      &   0.73$\pm$0.02 & 0.63$\pm$0.02  &  0.36$\pm$0.01    \\
                &   1400      &   1001   &   138  &   76 & 88  &   38.0$\pm$0.4 &   48.0$\pm$0.6  &   0.500$\pm$0.003 & 0.434$\pm$0.003 &  0.2762$\pm$0.002\\
                &   730      &   1975   &   233  &   131 & 151  &   48$\pm$7     &   61$\pm$9      &   0.37$\pm$0.05     & 0.32$\pm$0.05  &  0.21$\pm$0.03    \\
J1022$+$1001    &   3100      &   1626   &   371  &   169 & 208 &   130$\pm$70   &   280$\pm$140   &   0.8$\pm$0.5     & 0.6$\pm$0.4  & 0.4$\pm$0.2      \\
                &   1400      &   1963   &   969  &   122 & 150  &   134$\pm$6    &   290$\pm$15    &   1.10$\pm$0.04   & 0.89$\pm$0.04 &  0.138$\pm$0.006  \\
                &   730      &   1873   &   823  &   175 & 215 &   80$\pm$30    &   70$\pm$13     &   0.5$\pm$0.1     & 0.4$\pm$0.1  &  0.10$\pm$0.04    \\
J1603$-$7202    &   3100      &   1579   &   287  &   149 & 192 &   $<$277       &   $<$560        &   $<$2            & $<$1  &  $<$0.8           \\
                &   1400      &   1723   &   1210 &   147 & 190 &   146$\pm$31   &   300$\pm$56    &   1.0$\pm$0.2     & 0.8$\pm$0.2  &  0.12$\pm$0.03  \\
                &   730      &   2020   &   1342 &   216 & 279  &   $<$281       &   $<$570        &   $<$1            & $<$1  &  $<$0.2           \\
J1713$+$0747    &   3100      &   380    &   107  &   59  & 64  &   37$\pm$12    &   40$\pm$10     &   0.6$\pm$0.2     & 0.6$\pm$0.2  &  0.3$\pm$0.1      \\
                &   1400      &   377    &   108  &   63  & 68  &   31.1$\pm$0.7 &   35.0$\pm$0.8  &   0.49$\pm$0.01   & 0.46$\pm$0.01 &  0.29$\pm$0.01  \\
                &   730      &   568    &   221  &   126 & 136  &   $<$174       &   $<$200        &   $<$1            & $<$1  &  $<$0.8           \\
J1744$-$1134    &   3100      &   224    &   97   &   65  & 75  &   $<$171       &   $<$200        &   $<$3            & $<$2  &  $<$2             \\
                &   1400      &   245    &   138  &   73  & 84  &   35.5$\pm$0.7 &   37.8$\pm$0.8  &   0.49$\pm$0.01   & 0.423$\pm$0.009 &  0.258$\pm$0.006  \\
                &   730      &   273    &   150  &   73  & 84  &   $<$161       &   $<$200        &   $<$2            & $<$2  &  $<$1             \\
J1909$-$3744    &   3100      &   74     &   35   &   19  & 21   &   $<$34        &   $<$300        &   $<$2              & $<$2 &  $<$1             \\
                &   1400      &   88     &   42   &   24  & 26  &   10$\pm$1  &   8.6$\pm$0.8   &   0.40$\pm$0.04   & 0.36$\pm$0.04  &  0.23$\pm$0.03  \\
                &   730      &   110    &   60   &   28  & 31  &   $<$61        &   $<$60         &   $<$2            & $<$2  &  $<$1             \\
J1939$+$2134 & 3100  & 859 & 46 & 21 & 23 & $<$ 73  & $<$48 &  $<3.5$  & $<3.2$ & $<0.08$ \\
 	       &1400 & 854 & 12 & 16 & 17 & 5$\pm$1 & 6$\pm$1  &  $0.3\pm0.09$ & $0.3\pm0.1$ & $0.007 \pm0.002$  \\
	      & 730  &   863 &  820 &  17 & 19 & $19 \pm 1$ & $13 \pm 1$ & $1.15\pm 0.06$ & $1.03 \pm0.6$ &  $0.023 \pm 0.001$\\
J2145$-$0750    &   3100      &   4090   &   350  &   199 & 217 &   200$\pm$48   &   420$\pm$99    &   1.0$\pm$0.3     & 0.9$\pm$0.3  &  0.6$\pm$0.1      \\
                &   1400      &   4170   &   340  &   205 & 224 &   91$\pm$3     &   192$\pm$6     &   0.44$\pm$0.02 & 0.407$\pm$0.01 &  0.267$\pm$0.009  \\
                &   730      &   4194   &   413  &   204 & 223 &   80$\pm$15    &   170$\pm$38    &   0.40$\pm$0.08   & 0.37$\pm$0.06 &  0.20$\pm$0.04    \\
\hline
J0613$-$0200    &      1400         &   925    &   465  &   43 & 47 &   $<$409       &   $<$400             &   $<$9  & $<$9      &  $<$0.9 \\
J0711$-$6830    &      1400         &   2561   &   1899 &   73 & 80 &   $<$71        &   $<$90            &   $<$1 & $<$0.9    &  $<$0.04\\
  J1017$-$7156               &   1400      &   142    &   72   &   36  & 39   &    $< 124$ & $<100$        &  $<3.4$ & $<3.2$     & $<0.9$  \\
J1024$-$0719    &      1400         &   1464   &   497  &   65 & 71 &   $<$255       &   $<$300             &   $<$4& $<$4      &  $<$0.5 \\
J1045$-$4509    &      1400         &   1445   &   756  &   274 & 299 &   $<$623       &   $<$900             &   $<$2& $<$2      &  $<$0.8 \\
J1600$-$3053    &      1400         &   407    &   93   &   61 & 67 &   $<$234       &   $<$200             &   $<$4  & $<$4      &  $<$2   \\
J1643$-$1224    &      1400         &   928    &   319  &   206 & 225 &   $<$402       &   $<$500             &   $<$2  & $<$2      &  $<$1   \\
J1730$-$2304    &      1400         &   1712   &   976  &   97 & 106 &   $<$258       &   $<$400             &   $<$3& $<$2     &  $<$0.3 \\
J1732$-$5049    &      1400         &   1617   &   295  &   171 & 186 &   $<$659       &   $<$800             &   $<$4  & $<$4      &  $<$2   \\
J1824$-$2452A   &      1400         &   1600   &   979  &   30 & 33 &   $<$277       &   $<$300             &   $<$9& $<$8      &  $<$0.2 \\
J1857$+$0943    &      1400         &   3011   &   523  &   103 & 112 &   $<$219       &   $<$300             &   $<$2& $<$2      &  $<$0.4 \\
J1939$+$2134    &      1400         &   859    &   65   &   25 & 27 &   $<$15        &   $<$10            &   $<$0.6 & $<$0.5   &  $<$0.2 \\
J2124$-$3358    &      1400         &   876    &   124  &   140& 153 &   $<$341       &   $<$400             &   $<$2& $<$2      &  $<$3   \\
J2129$-$5721    &      1400         &   617    &   265  &   75 & 82 &   $<$386       &   $<$400             &   $<$5  & $<$5      &  $<$1   \\
J2241$-$5236    &      1400         &   123    &   64   &   28  & 31 &   $<$58        &   $<$50            &   $<$2     & $<$2      &  $<$0.9 \\
\hline
\end{tabular}\\
\end{center}
 Notes:  For each pulsar we list measurements at different observing frequencies.  We list the pulse width measured at $10\%$ of peak intensity $W_{\rm 10}$, $50\%$ of peak intensity ($W_{\rm 50}$), and the effective pulse width $W_{\rm eff}$, as defined in Equation (\ref{eqn:effective_width_sc}) .    We also list the rms noise expected due to jitter per pulse ($\sigma_J(1))$, and expected in an hour-long observation $\sigma_J({\rm hr})$.  We calculate the jitter parameter using, $W_{\rm eff}$, $W_{ 50}$, and $W_{\rm eff}(1+m_I^2)$ as proxies for pulse width.   For the pulsars for which we did not measure the modulation index $m_I$, we assumed $m_I =0.3$. 
\end{table*}

\section{Improving pulsar timing in the presence of jitter noise} \label{sec:improve}

 \subsection{Including  jitter noise in timing models} \label{sec:ejit}

In pulsar timing observations it is common to account for unknown uncertainties by  1) multiplying the TOA uncertainty  by an arbitrary factor (EFAC); 2) adding to the TOA uncertainty an additional  term in quadrature (EQUAD); or both \cite[][]{2006MNRAS.372.1549E}.   

EFACs were originally included in the pulsar timing program \textsc{tempo} to account for the fact that the reduced-$\chi^2$ of the best fitting models were typically greater than unity.
EQUAD factors were
included to avoid over-emphasising high-S/N observations in weighted fits.
Over-weighted points effectively reduce the number of degrees of freedom
for the fit and, in the presence of systematic TOA errors, can easily bias
the timing solution.


 Common EQUAD or EFAC values  are typically applied to all the TOAs in a pulsar timing dataset, or to large subsets that are expected to have identical values, such as  TOAs derived from the same backend instrument at the same observing radio frequency.   
 Various methods have been used to estimate their values.  
The most common method is to  adjust the values until the reduced-$\chi^2$ of the fitted model reaches unity.
 Bayesian and other  maximum likelihood methods have also recently been developed and applied to precision pulsar timing datasets \cite[][]{2009MNRAS.395.1005V,2014MNRAS.437.3004L}.  In these methods, EFAC and EQUAD are included as nuisance parameters and  marginalised when calculating the posterior distributions of parameters of interest or comparing models. 

Instead of modelling the factors EQUAD and EFAC from the timing datasets, we   add a term associated with pulse-shape variations, derived from our single-pulse and intra-observation analysis, and not as part of the timing model. 
We modify the TOA uncertainty  $\Delta_{\rm TOA}$ by adding a term to account for pulse jitter:
\be
\label{eqn:new_error}
\Delta_{\rm TOA}^2 = \Delta_F^2 + \left(\frac{  \sigma_J }{\sqrt{N_p}}\right)^2= \Delta_F^2 +  \sigma^2_J(1)  \frac{P}{T},
\ee
where $N_p = T/P$ is the number of co-added pulses, $P$ is the pulse spin period, $T$ is the observing span, $\Delta_F$ is the formal TOA uncertainty,  and $\sigma_J(1)$ is the level of jitter noise expected from a single pulse.  

As an example, we analyse a subset of observations for PSR J0437$-$4715, the brightest MSP in our sample, and one where we expect all of our observations to show evidence for pulse jitter. 
  We use observations conducted in the $10$cm band with the PDFB4 backend taken from the end of its commissioning in $2008$ to the end of  $2013$ (MJDs 54753 to  56646).
  We corrected the observations for dispersion measure variations using multi-frequency PPTA data and a technique that has been shown not to remove noise from pulsar timing residuals \cite[][]{2013MNRAS.429.2161K}. 
  In contrast to the PPTA analysis,  in which pulse profiles are formed from the invariant interval  \cite[][]{2000ASPC..202...73B},  we formed profiles from  total intensity (Stokes $I$ parameter).
      Observations varied from approximately $2$~min to $64$~min in duration, enabling us to compare the correction scheme given in  Equation (\ref{eqn:new_error}) to established techniques.
      
\begin{figure}
\begin{center}
\includegraphics[scale=0.5]{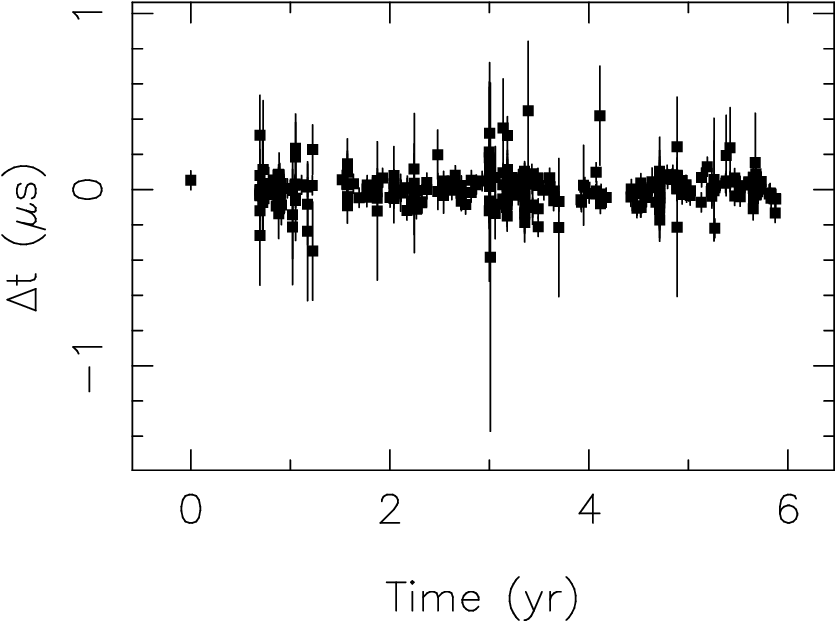}
\caption{  \label{fig:resid_ejit_J0437}  Residual TOAs for PSR J0437$-$4715. The TOA uncertainties include a term to account for jitter noise.    \footnotesize        }  
\end{center}
\end{figure}

    After modifying the TOA uncertainties,   the timing model presented in \cite{2013PASA...30...17M} was refitted.  
     In Figure \ref{fig:resid_ejit_J0437}, we plot the residual TOAs of this  best-fitting model.   
We find that the weighted rms of the residuals over $5$ years is $64$~ns and the reduced~$\chi^2$ of this model is $1.1$.   

\begin{figure}
\begin{center} 
\includegraphics[scale=0.4]{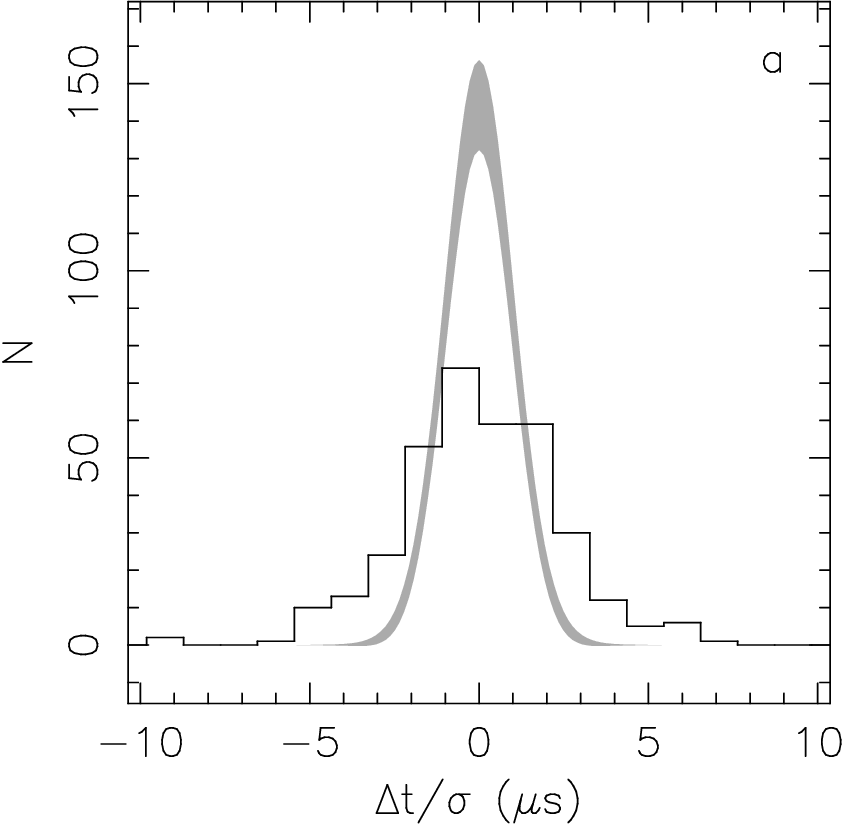} \\
\vspace{2mm}
\includegraphics[scale=0.4]{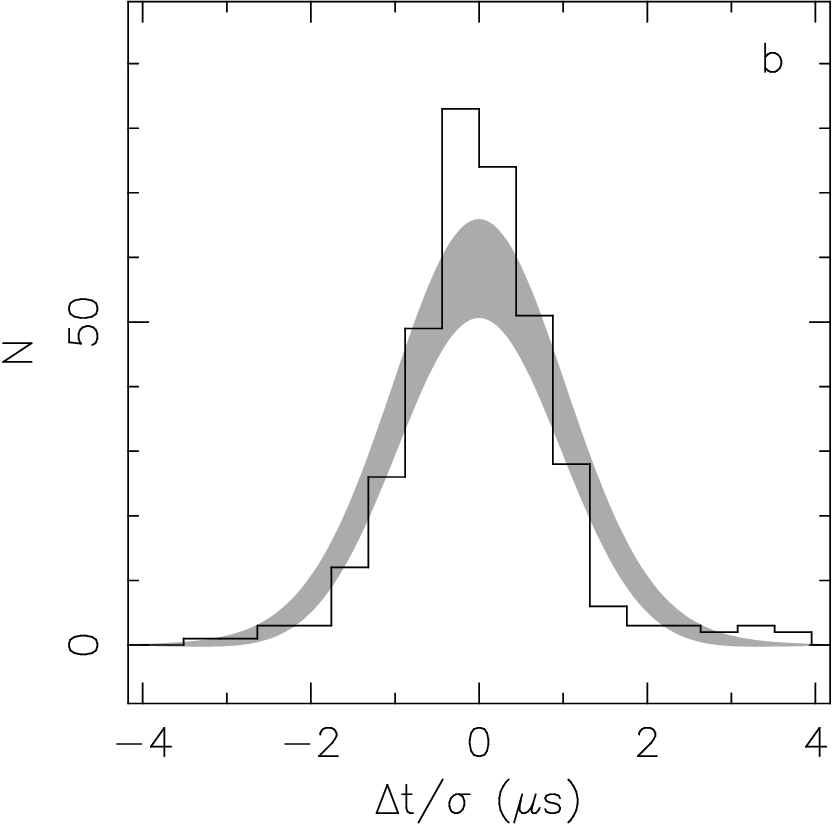}  \\
\vspace{2mm}
\includegraphics[scale=0.4]{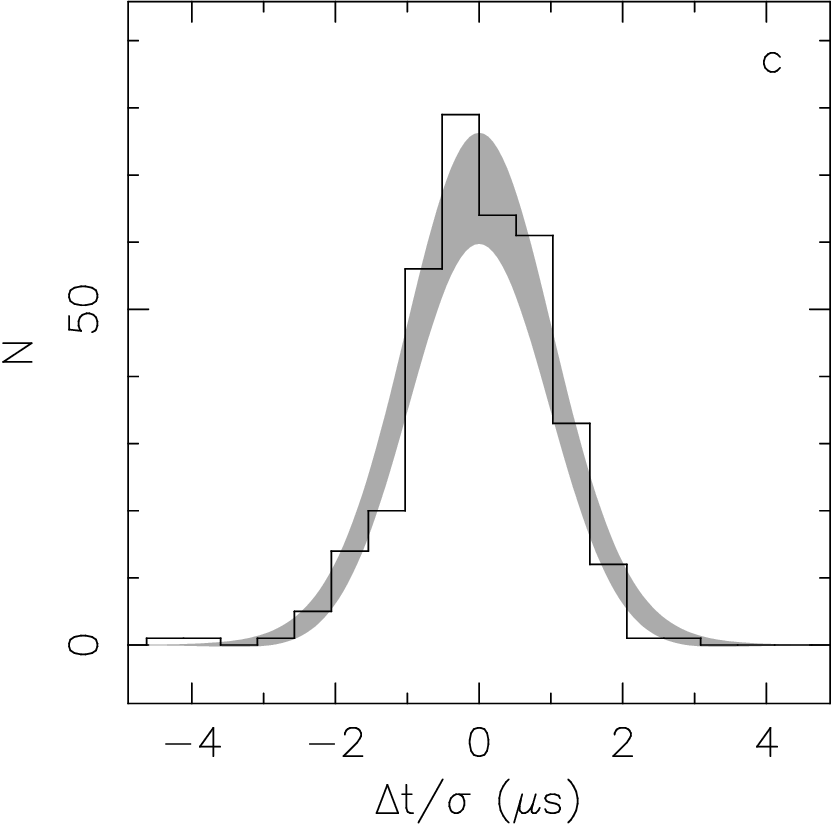} \\
\caption{  \label{fig:ejit_J0437}  \footnotesize   Histograms of normalised residuals for PSR~J0437$-$4715.  Panel a):   No modification to TOA uncertainties  Panel b):  Adding in quadrature $100$~ns to the TOA uncertainties.   Panel c):   Modifying TOA error using Equation (\ref{eqn:new_error}).   The grey shaded area represents the range in the histogram theoretically expected.         }  
\end{center}
\end{figure}

We compared the results of this noise model   to two alternate cases that have previously been applied to pulsar timing datasets:    
a) fitting is conducted with  no modification  to the TOA uncertainties and, as discussed above, b) including an  EQUAD term  such that  the  reduced-$\chi^2$ of the residuals for the best-fitting model is unity.
             In Figure \ref{fig:ejit_J0437}, we show histograms of the normalised residuals from the three fits, and the  normal  distribution expected if the uncertainties match the arrival times.  
In case   (a),  after fitting,  we find that the reduced $\chi^2$ was approximately $8$, and the distribution of uncertainties is much broader than expected.  
    In case (b), the added EQUAD factor (panel b)  gave a reduced $\chi^2$ of unity; however, the  distribution of uncertainties is too narrow.
 The uncertainties of the high-precision long-integration TOAs have been suppressed to account for outlying short observations present in the observations.   
    Including jitter explicitly  (panel c) we find the that the normalised  TOAs well match the expected distribution.  
 The distribution of the TOA uncertainties  best matches a Gaussian distribution. 

In order to assess the improvement of our weighting schemes to cases a) and b), we produced power spectra of the residuals.
In all cases, we found that the power spectra were white.
To compare the methods, we measured the amplitude of the power spectral density (PSD).
The amplitude of the PSD in our weighting scheme is  $\sim 20\%$ lower than case a) and $\sim 27\%$ lower than case b).
This suggests that TOAs weighted using our scheme are more sensitive to the GWB.

    We emphasise that we have corrected the error bars using a physically motivated technique that uses an {\em a priori} model of the residuals.    
In addition to providing a better model of the true TOA uncertainties,  EQUAD and EFAC parameters may not need need to be modelled in timing datasets.
This reduces by two  the dimensionality to TOA modelling, which streamline computationally intensive Bayesian GW-search algorithms \cite[][]{2009MNRAS.395.1005V}.   In archival data, it may be necessary to include EFAC or EQUAD parameters to account for pulse shape distortions induced by non-linearities in instrumentation associated with low-bit digitisation.

\subsection{Timing a sub-selection of pulses}

For many pulsars, the brightest pulses originate in a narrow region of pulse phase.
Examples of this is are the giant pulses  from the MSP PSR~J1939$+$2134, which are emitted in only a narrow window of pulse phase. 
This suggests that it may be possible to form a more precise TOA by producing a pulse profile from a selection of pulses.   
Selecting bright pulses comes at a sacrifice of averaging fewer pulses together, exacerbating the effects of any jitter present in the subset.    
If instead of timing all of the pulses, a fraction $f$ of the $N_p$ pulses is used  the expected timing precision would be
\be
\label{eqn:rms_frac_pulses}
\sigma_J (f, N_p)  =  \frac{\sigma_J(f, 1)}{\sqrt{f N_p}}, 
\ee
where $\sigma_1(f, N_p)$ is the scatter in the $f N_p$  selected single pulses.

From Equation (\ref{eqn:rms_frac_pulses}) we can derive a condition on the timing improvement achieved by using a fraction of the pulses. Improvement is  achieved if and only if 
 \be
 \label{eqn:jit_improve}
 \sigma_J(f, 1)   <   \sigma_J(1,1) \sqrt{f}.
 \ee
The brightest pulses must originate in a narrow region of pulse phase to achieve an improvement in timing precision.

In order to see if an improvement could be obtained, we tested Equation (\ref{eqn:jit_improve}) using our fast-sampled observations of PSRs J0437$-$4715 and PSR~J1909$-$3744.
We formed profiles from only fractions $f$  of the brightest pulses.  We then crosscorrelated these profiles with the standard template, and measured the resulting rms of the corresponding time series.     

\begin{figure}
\begin{center} 
\includegraphics[scale=0.5]{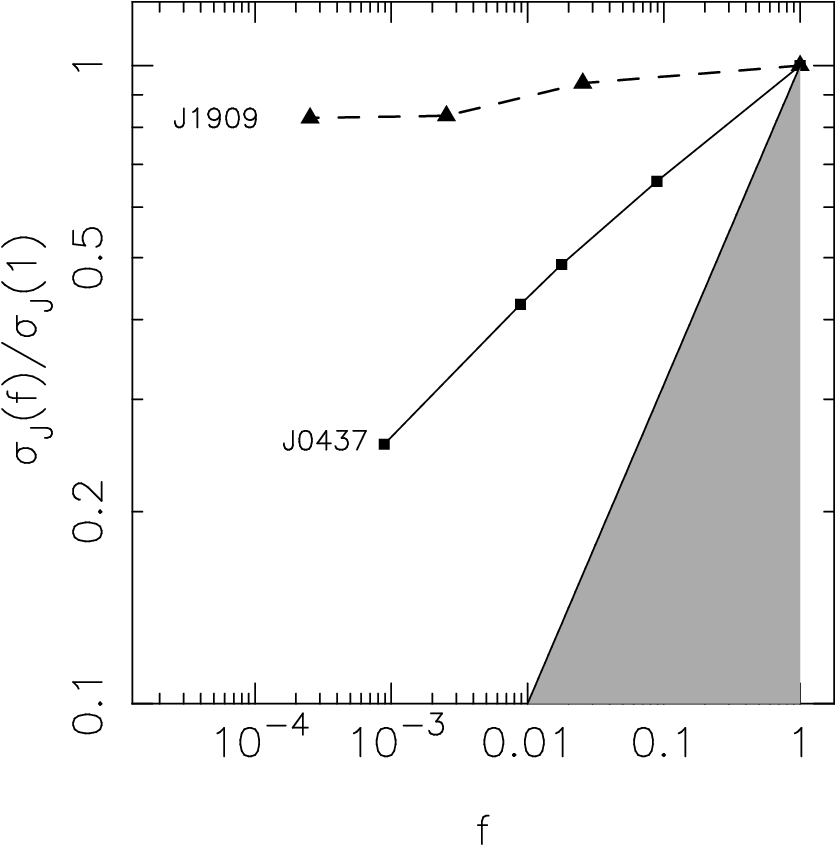}
\caption{  \label{fig:bright_sn}  \footnotesize  Fraction of pulses used versus normalised timing error  for PSR J0437$-$4715 (labelled J0437)  and PSR~J1909$-$3744 (labelled J1909).   The filled grey area identifies the region $\sigma_J(1,f) < \sigma_J(1,1) / \sqrt{f}$,    in which precision timing would benefit from using fewer pulses.      }  
\end{center}
\end{figure}

In Figure \ref{fig:bright_sn} we  see how $\sigma_J$ depends on $f$ for these pulsars.
For PSR J0437$-$4715, the rms error does decrease when selecting only the brightest pulses as the brightest pulses originate from a narrow region of pulse phase.  
 However, it does not decrease sufficiently quickly to warrant timing only the brightest pulses and better timing precision is achieved by timing all of the pulses.

PSR J1909$-$3744 shows only a modest decrease in the timing error when bright pulses are selected.   The rms error in the subselection only reduces to  a level of $\approx 80\%$ of the rms error of all of the pulses.
 This suggests  that bright pulses span a region as nearly as wide 
 as   the  entire main pulse, consistent with the profile of the brightest pulses  (see Figure \ref{fig:energy_hist_J1909}).  

\cite{2014oslo} tested both this idea and a related idea of timing only the weakest pulses.  In this manner the effects of pulse jitter can be minimised and a better $\chi^2$ of the timing model can be achieved when including only data with instantaneous S/N of less than unity. This yields more realistic formal TOA uncertainties.  However using Equation (\ref{eqn:new_error}) we can now account for pulse jitter and correct the formal TOA uncertainties.

\section{Discussion and conclusions} \label{sec:discuss}

In the highest S/N observations of the  MSPs in our sample, we find that there is a contribution to timing uncertainty in excess of what is expected from radiometer noise alone.   We attribute this contribution to stochastic shape variations. 
For PSR~J1713$+$0747, we find an excess that is common to  multiple instruments and different telescopes. 
  Our results suggest  pulse  jitter is a generic property of MSPs (and pulsars in general) and that all MSPs will show jitter if observed with sufficient sensitivity.  This is to be expected because the pulsar emission can be thought of as a noise source \cite[][]{1975ApJ...197..185R,2011MNRAS.418.1258O,2013MNRAS.430..416O}.
We estimate that the   rms contribution of jitter is   $ \sigma_J(T) \approx 0.5  W_{\rm eff} \sqrt{P/T}$ where $P$ is the pulse spin period, $T$ is the observing span and $W_{\rm eff}$ is the effective width defined in Equation \ref{eqn:effective_width_sc}.
Because pulse widths typically decrease with higher observing frequencies,  we find that jitter  noise  is lower in the 10cm band than in the 20cm or 50cm bands.

We find that we can better account for TOA uncertainty when we include in quadrature a jitter noise term extrapolated from our single-pulse estimates.  If jitter noise is not properly included in the timing model (either being ignored or through an erroneously applied EQUAD) the longer lower-gain observations from smaller telescopes will be improperly down-weighted.

Consistent with previous observations, we find that jitter noise is correlated within observing bands.
For PSR~J0437$-$4715 there is sufficient S/N to detect jitter noise simultaneously at both $730$~MHz and $3100$~MHz in the dual band 10cm/50cm receiver.
While jitter noise is present in both bands, it is uncorrelated between the bands.  
  Either the pulse emission decorrelates between the bands or the components  contributing the jitter noise are different.
Over wide bandwidths pulsar emission has varying levels of correlation, with some pulsars showing high levels of correlation in pulse energy over $3$~GHz bandwidth  \cite[][]{1978A&A....70..307B}, others showing little correlation  between $0.08$ and $1.4$~GHz   \cite[][]{1968Natur.218.1143R}, and others showing intermediate behaviour \cite[][]{2007A&A...462..257B}.
Dual-band single-pulse datasets for PSR~J0437$-$4715 would enable better characterisation of the nature of the correlation of the pulse shape variations.
For the other pulsars in the sample, the dual frequency analysis was not possible because they do not have the sensitivity to detect jitter noise simultaneously in both bands of the 10cm/50cm receiver. Such analysis would be feasible with more sensitive telescopes, if they are equipped with dual-band or broad-band receiving systems.

If jitter noise decorrelates over finite bandwidth, it is possible to mitigate its effects with sufficiently wide-band instrumentation.
Many future receiving systems will have   wide bandwidths, sampling from a frequency range of $\sim 0.5 - 4$~GHz of great utility for pulsar timing.
These systems may sample a few independent realisations of jitter, reducing its rms contribution to overall timing uncertainty. 
However the decorrelation may introduce a bias if a single-epoch dispersion variation correction strategy is employed \cite[][]{2013ApJ...762...94D} because the frequency dependent shape variations may be interpreted as dispersion measure variations \cite[][]{2014arXiv1402.1672P}.

We find that most of the MSPs for which we could detect single pulses  show log-normal pulse-energy distributions. 
Our pulse-energy statistics show similarities to slower pulsars.  In a sample of $\approx 350$ pulsars \cite{2012MNRAS.423.1351B} found that the majority showed log-normal or nearly log-normal energy distributions, with a minority showing Gaussian energy distributions. 
We find no evidence for power-law pulse-energy distributions or giant pulses in our single-pulse observations.
The pulsar that has the lowest level of jitter noise, PSR J1909$-$3744, does not  show evidence for a log-normal tail in pulse energies, which results in a  relatively low level of jitter noise.  
\cite{2004MNRAS.353..270C}  found that  in the slowly spinning pulsars in their study, the edges of the pulse profiles show Gaussian energy modulation.    
The edges of the pulse were interpreted as arising from the edge of the open field line region.
This suggests, that for PSR J1909-3744, our line of sight may
traverse the edge of an emission region.

We have not explored algorithms for correcting datasets for  pulse-shape variability. In order for  this to work it is necessary to find a strong correlation between the pulse shape and another measurable quantity of interest, because the error in the correction will be $\propto 1-\rho^2$ , where $\rho$ is the correlation coefficient \cite[][]{cs2010}.
    \cite{2011MNRAS.418.1258O,2013MNRAS.430..416O} have identified methods to correct for pulse shape variability in PSR~J0437$-$4715, reducing the rms of the residuals by $20\%$ to $40\%$.   \cite{2012ApJ...761...64S}  identified only a weak correlation between pulse arrival time and S/N in PSR~J1713$+$0747. The level of correlation was insufficient to implement a correction scheme.   

We also have not explored optimal intra-observation weighting schemes.  Typically pulse profiles are formed, after removing RFI and calibrating, combining the many sub-integrations.
  These sub-integrations are typically combined using natural weighting, with the higher S/N portions of the observations given greater weight. 
If a pulsar is in a jitter-dominated state (at least in the brightest observations), this  scheme over-weights the brightest portion of the observation. 
This reduces the effective number of pulses in the profile. 
 To mitigate this effect TOAs could be produced  from shorter sub-integrations  and the errors corrected using Equation \ref{eqn:new_error}.   Another possibility would be to combine the sub-integration using a weighting scheme that accounts for the presence of S/N independent noise.

These results are relevant to PTA activities on current and future telescopes.  
It is important to incorporate the effects of jitter when considering the sensitivity of pulsar timing array experiments to gravitational waves \cite[][]{2012ApJ...750...89C}, and to optimise observing strategies.
Based on our results, we expect that pulse-shape variations limit the timing precision at the larger aperture telescopes that are part of the North American Nanohertz Observatory for Gravitational Waves \cite[NANOGrav, ][]{2013ApJ...762...94D} and the European Pulsar Timing Array \cite[EPTA,][]{2010CQGra..27h4014F}.
   Observations of bright MSPs made using very long baseline interferometry with the   Large European Array for Pulsars \cite[LEAP, ][]{2013CQGra..30v4009K}, which has a sensitivity comparable to Arecibo, will also be limited by jitter noise.  Including an improved noise model is especially important for producing datasets from observations with different integration times, such as for the  International Pulsar Timing Array project \cite[][]{2010CQGra..27h4013H},  which combines EPTA, NANOGrav, and PPTA data.
    The IPTA datasets     contain   short-integration high-sensitivity TOAs from large-aperture telescopes  that   need to be properly combined with less sensitive longer integration observations.
As such, if the effects are not incorporated in the timing analysis, a modest loss in sensitivity to the GWB would be expected.
The IPTA project could also consider alternative scheduling strategies in which the smaller-aperture telescopes focus on the bright pulsars that would be jitter-dominated  when observed at larger-aperture telescopes. The larger telescopes could then then focus on fainter pulsars. 
 

The IPTA pulsars will all be candidates  for  PTA observations with MeerKAT, the Five Hundred Metre Aperture Spherical Telescope (FAST),  and the Square Kilometre Array (SKA).  These telescopes will all have larger collecting area than the Parkes telescope.
Observations with these telescopes are likely to be jitter-noise limited for many of these pulsars.
Sub-array modes in which the array can be split and observe multiple pulsars independently would maximise the timing throughput.  
Alternatively, in the era of the SKA, the brighter timing-array pulsars could be observed with $100$-metre class radio telescopes. 

\section*{Acknowledgments}

We thank the referee for helpful comments that improved the clarity of the text. The Parkes radio telescope is part of the Australia Telescope National Facility which is funded by the Commonwealth of Australia for operation as a National Facility managed by CSIRO.  This work was performed on the gSTAR national facility at Swinburne University of Technology. gSTAR is funded by Swinburne and the Australian Government?s Education Investment Fund.     This work was supported by the Australian Research Council through grant DP140102578.   GH  is a recipient  of a  Future Fellowship from the Australian Research Council.    VR is a recipient of a John Stocker postgraduate scholarship from the Science and Industry Endowment Fund of Australia.  LW acknowledges support from the Australian Research Council.  This work made use of NASA's ADS system.  

\end{document}